%% file: main.tex
\documentclass{article}

\usepackage{arxiv}
\usepackage{amsfonts}       
\usepackage{xcolor}

\usepackage[authoryear]{natbib}
\usepackage{hyperref}

\usepackage{paralist}

\usepackage{setspace}
\usepackage{amsmath}
\usepackage{amssymb}
\usepackage{mathtools}
\usepackage{mismath}
\usepackage{derivative}
\usepackage{subcaption}
\usepackage{lineno}

\usepackage{float}
\usepackage{booktabs}
\usepackage{paralist}
\usepackage{array}
\usepackage{algpseudocode, algorithm, algorithmicx}

\newcommand{\Idiffx}{\, {\rm d\vect{x}}}
\newcommand{\Idiffr}{\, {\rm d\vect{r}}}

\newcommand{\boldparagraph}[1]{\paragraph{\textbf{#1}}}

\title{Numerical Investigation of Elastically-Mounted tandem Cylinders using an ALE Runge-Kutta Discontinuous Galerkin method}

\author{
    Alexios Papadimitriou \\
    School of Naval Architecture and Marine Engineering\\
    National Technical University of Athens\\
    Zografos, 15780, Athens \\
    \And
    Spyridon Zafeiris \\
    School of Naval Architecture and Marine Engineering\\
    National Technical University of Athens\\
    Zografos, 15780, Athens \\
    \And
    George Papadakis \\
    School of Naval Architecture and Marine Engineering\\
    National Technical University of Athens\\
    Zografos, 15780, Athens \\
}

\begin{document}
    \boldvect
    \maketitle

\begin{abstract}
   This work presents a high-order Arbitrary-Lagrangian-Eulerian (ALE) Discontinuous Galerkin framework for simulating multi-body Vortex-Induced Vibrations. The ALE formulation extends a Runge-Kutta Interior-Penalty nodal DG solver with minimal additional computational overhead, incorporating discrete enforcement of the Geometric Conservation Law (GCL) to ensure free-stream preservation and Radial Basis Function (RBF) mesh deformation to handle large structural displacements. The framework is applied to elastically-mounted tandem cylinder configurations: a two-cylinder arrangement with cross-flow oscillations at Re=200, and a three-cylinder arrangement with two degrees of freedom at Re=150. In the three-cylinder case, the trajectories exhibit highly irregular behavior driven by complex wake interference, including a periodic attract-and-release mechanism governing the trailing cylinder's stream-wise response. Results are verified against established benchmarks through Lissajous curves, Poincar\'{e} phase maps, power spectra, and vortex shedding mode classification. An hp-refinement comparison demonstrates that increasing the polynomial order is more effective and computationally efficient than mesh refinement for capturing multi-body wake dynamics, as the low numerical diffusion of the high-order method preserves vortical structures over long distances on relatively coarse meshes. These findings highlight the importance of high-order methods for CFD-FSI applications where wake interactions drive the structural response.
\end{abstract}

\maketitle

\section{Introduction}\label{intro}
\input{intro2}

\section{Methodology} \label{methodology}
\input{methodology}

\section{Numerical Results} \label{numerical_results}

\input{numerical_results}

\section{Conclusions}\label{conc}

\input{conclusions3}

\section*{Acknowledgments}
We acknowledge EuroHPC JU for awarding the project ID EHPC-REG-2025R01-095 access to Karolina, which was used for producing this work's results.

\section*{Funding}
The research work was supported by the Hellenic Foundation for Research and Innovation (HFRI) under the 5th Call for HFRI PhD Fellowships (Fellowship Number: 20716), which involves the second author.









\appendix
\renewcommand{\thesection}{Appendix}

\section{On the GCL equation}
\label{sec:appendix-gcl}

\setcounter{equation}{0}
\renewcommand{\theequation}{A\arabic{equation}}

\input{appendix}


\bibliographystyle{plain}


\bibliography{cas-refs}



\end{document}

%% file: intro2.tex
Fluid-Structure Interaction (FSI) problems characterized by large relative boundary motions pose a major challenge in Computational Fluid Dynamics (CFD). Achieving high-fidelity resolution of the complex, non-linear flow features inherent to these problems, such as separation, wake interference, and structural resonance, demands advanced numerical discretization strategies \cite{bungartz2006fluid}.  In traditional low-order methods, maintaining a fine mesh resolution throughout the extended wake region leads to prohibitively large grids; see, for example, the work of Gao et al. \cite{gao2020numerical},
whereas gradual mesh coarsening introduces excessive numerical diffusion that artificially dampens the vortical structures. 
To mitigate numerical diffusion without incurring massive grid sizes, high-order methodologies are effective candidates \cite{wang2007high}. Among these, the \textit{Runge-Kutta} time-marched \textit{Discontinuous Galerkin} (RK DG) method \cite{cockburn1998tvbV} has emerged as a promising option. By providing arbitrary order of spatial accuracy on unstructured grids, the RK DG method remedies excessive numerical diffusion and is, therefore, ideally suited for resolving fine-scale vortical structures. However, despite the DG method's growing popularity for addressing complex flow problems \cite{wang2013discontinuous,lomtev1999discontinuous,fidkowski2016hybridized}, its extension to moving domains and application to multi-body FSI cases remains sparse in the literature.

A key challenge in multi-body FSI is accommodating the relative body motion. To this end, solvers typically employ either overset (also known as ``chimera'') grids \cite{wurst2015,galbraith2015,brazell2016}
or Arbitrary-Lagrangian-Eulerian (ALE) frameworks. While overset grids offer kinematic flexibility for large displacements, they introduce interpolation errors, potential degradation of numerical accuracy, and severe algorithmic and performance bottlenecks during parallel execution, see e.g. \cite{volkner2017analysis,zafeiris2024overset}. In addition to ``chimera'' grids, \textit{Adaptive Mesh Refinement}, where the mesh is locally refined as the body moves, has been applied in the context of VIV using a DG method by Zou et al. \cite{zou2025moving}, who demonstrated accurate single-cylinder VIV predictions. Conversely, ALE frameworks deform a single continuous mesh in response to boundary motion, inherently favoring conservation.

While DG methods have been proven advantageous for static meshes, for example \cite{Bolemann2015}, coupling them with an ALE framework introduces extra algorithmic steps together with certain computational challenges. For instance, mesh deformation requires the re-evaluation of all DG matrices at every time step. This overhead is alleviated here by using affine (straight-sided) triangles and a mapping to a reference space so that only the element's constant Jacobian matrix needs to be updated. Additionally, the full-discrete system must account for the grid velocity, which is typically achieved either by evaluating the fluxes directly in the moving physical space or by mapping the equations back to a static reference element. Another requirement for ALE formulations is the strict satisfaction of the \textit{Geometric Conservation Law} (GCL), which dictates the solver's ability to enforce conservation, and is usually achieved by enforcing conservation for a uniform free-stream flow. While ALE-DG evaluated directly on physical space inherently satisfies the GCL, as shown by Nguyen \cite{nguyen2010arbitrary}, the standard practice of mapping to a reference element introduces numerical integration errors that require careful treatment to ensure discrete conservation; we refer the reader to \cite{persson2009discontinuous}.

Having outlined the main algorithmic challenges, we now turn to the physical problem used to evaluate the numerical framework. Investigating \textit{Vortex Induced Vibrations} (VIV) serves as a particularly demanding test case from both the numerical and physical perspectives. VIV occurs when alternating vortices shed by bluff bodies induce structural oscillations. This is a non-linear interaction that governs the design of various engineering systems, including marine risers, subsea cables \cite{wen2025modal,trim2005experimental,hover1997vortex}, and aquatic clean energy converters like VIVACE \cite{bernitsas2008vivace}. In particular, arrays of closely-spaced marine risers and bundled pipeline configurations, commonly encountered in offshore oil and gas operations, are subject to complex wake interference that can amplify fatigue loads well beyond what isolated cylinder predictions would suggest \cite{gao2020numerical,xu2021tandem}. Beyond its engineering relevance, VIV also highlights the need for high-order methods. Because VIV is driven by flow separation and complex wake-body interference, the resulting body responses are sensitive to numerical diffusion. Therefore, accurately capturing the wake evolution and its interaction with the moving bodies can have a significant impact on the predicted response.

Given the complexity of engineering VIV problems,  simplified systems are usually studied in the literature. Early VIV studies primarily focused on a single cylinder with one \textit{Degree of Freedom} (DoF) oscillating in the cross-flow direction \cite{williamson2004vortex}. For low mass and damping ratios, these studies identified distinct response branches and synchronization phenomena \cite{khalak1996dynamics,brika1993vortex}, revealing specific vortex shedding modes such as the $2S$ (two single vortices) and $2P$ (two pairs) patterns. 
In \cite{konstantinidis2021resonance}, in-line oscillations were investigated. A single response branch was reported, characterized by small oscillation amplitudes of approximately $1\%$ of the cylinder's diameter. The response is associated with an $S$-mode vortex shedding pattern and resonance occurs when the vortex shedding synchronizes with the structural motion.

However, fluid-structure dynamics become vastly more complex for multi-body configurations, such as cylinders in tandem arrangements. For two tandem cylinders with one DoF, numerical studies predominantly relying on standard second-order \textit{finite-volume} (FV) discretizations \cite{xie2012numerical}, including those utilizing Immersed-Boundary \cite{griffith2017flow,borazjani2009vortex} and Hybrid Lagrangian-Eulerian solvers \cite{papadakis2022hybrid}, have shown that wake interference from the upstream cylinder dictates distinct oscillation branches and shedding modes for the downstream body.

When two DoF (cross-flow and in-line) are introduced, the kinematics become highly non-linear. While a single 2-DoF cylinder naturally follows a figure "8" trajectory \cite{jeon2001circular}, multi-body systems exhibit far more complex dynamics. For two cylinders with two DoF, experiments were carried out by Huang et al. \cite{huang2013experimental} showcasing these complex trajectories, while the effect of spacing was studied by Papaioannou et al. \cite{papaioannou2008effect}. The three-tandem-cylinder arrangement was studied numerically by Gao et al. \cite{gao2020numerical} and Yu et al. \cite{Yu2016FlowInduced} with the latter applying a solver discretized by a \textit{finite-element method} (FEM). In these multi-body systems, trajectories follow different patterns, including periodic, quasi-periodic, or even patterns without specific periodic characteristics. Experimental work on three tandem cylinders by Kang et al.\ \cite{kang2025experimental} has shown that both the spacing ratio and the number of cylinders significantly affect the downstream cylinder's response. Additionally, rigidly coupled cylinders in both tandem and parallel arrangements have been studied \cite{yin20242,gao2020flow}, revealing distinct vortex dynamics and wake topologies. Cylinders with non-circular profiles, particularly square sections, have also been investigated \cite{behara2024characterizing}; in such cases the angle of attack plays a significant role due to the asymmetry of the cross-section, leading to substantially different wake formations. Further arrangements, such as arrays of cylinders, see \cite{oviedo2013vortex}, or side-by-side configurations, see \cite{cui2014vortex, chen2019vortex, islam2023flow}, introduce parallel wake interactions. Notably, the  majority of these studies rely on low-order spatial discretization, and applications of high-order DG methods to multi-body VIV remain, is to the best of our knowledge, very limited.

In this work, an ALE nodal explicit Runge-Kutta Interior-Penalty Discontinuous Galerkin (RK IPDG) framework is developed and applied to several multi-body laminar VIV problems. The numerical solver discretizes the \textit{compressible Navier-Stokes} (cNS) system for low-Mach boundary conditions. Method-wise, this work verifies the statements of Persson et al. \cite{persson2009discontinuous} considering the GCL treatment, but on triangular meshes. As a result, the solver ensures conservation. Affine triangular elements are used here due to their memory-attractive properties, as we shall show in \autoref{methodology}. Also, our framework is presented without making assumptions about the polynomial base, albeit in practice nodal bases were used based on the original solver \cite{zafeiris2025parallel}. Within the ALE context, mesh deformation  is handled using \textit{Radial Basis Function} (RBF) interpolation \cite{rendall2009efficient,rendall2010reduced}. 
This way, high-order DG becomes computationally beneficial compared to low-order methods, since a small number of elements are deformed efficiently, each containing many polynomial DoFs.

A key finding of this study is that the low numerical diffusion of the high-order DG method preserves the vortical wake structures over long distances on relatively coarse meshes, which directly impacts the accuracy of the predicted multi-body VIV response. An $hp$-refinement comparison demonstrates that increasing the polynomial order is more effective and computationally efficient than mesh refinement for capturing the complex wake dynamics that drive the coupled cylinder motion.

To demonstrate the robustness and accuracy of the numerical framework, the remainder of this paper is structured as follows. Section~\ref{methodology} briefly describes the mathematical formulation of the ALE-DG method, outlining also the grid deformation strategy. Section~\ref{numerical_results} presents the numerical investigations, which focus on two-dimensional laminar flows. First, the importance of the GCL application at the discrete level is demonstrated. Subsequently, the 1-DoF two-cylinder tandem arrangement is investigated, comparing the resolved wake structures and response branches with the established data of Griffith et al. \cite{griffith2017flow} and Papadakis et al. \cite{papadakis2022hybrid}. In the final test case, a 2-DoF three-cylinder tandem configuration is examined, evaluating the performance of the DG solver against the literature data of Yu et al. \cite{Yu2016FlowInduced}. Section~\ref{conc} summarizes the conclusions. By investigating these complex multi-body configurations, this paper demonstrates the generalizability of the ALE-DG method, establishing a foundational benchmark for the application of the DG methodology in complex fluid-structure interactions.

%% file: methodology.tex
\boldparagraph{Notation}

In this section, the partial derivative with respect to $\alpha$ is symbolized as $\partial_{\alpha}  \equiv \partial / {\partial \alpha}$ and is applied component-wise to any array. The gradient of a vector field is $\nabla \vect{a} = \big[\, \partial_x \vect{a} \,| \, \partial_y \vect{a}\, |\, \partial_z \vect{a}\,\big]$
and the divergence of $\mathcal{A} = [\vect{a}_x\, | \, \vect{a}_y\, | \, \vect{a}_z]$, is $\nabla \cdot {\cal A} =  \partial_x \vect{a}_x + \partial_y \vect{a}_y + \partial_z \vect{a}_z$.

\subsection{The Compressible Navier-Stokes System}
The fluid is considered to be an ideal compressible gas with its dynamics being described by the compressible Navier-Stokes equations (cNS). These are written as the following system of \textit{Partial Differential Equations} (PDEs),
\begin{subequations}
	\begin{align}
		\partial_t \rho + \nabla \cdot \left( \rho \vect{u} \right)                                              & = 0 \label{eq:cont},                                                      \\
		\partial_t ( \rho \vect{u} ) + \nabla \cdot \left( \rho \vect{u} \otimes {\vect{u}} + p \text{I} \right) & = \nabla \cdot \tau \label{eq:momentum},                                  \\
		\partial_t \epsilon + \nabla \cdot \left( \epsilon \vect{u} + p \vect{u} \right)                         & = \nabla \cdot \left( \tau \vect{u} - \vect{q} \right) \label{eq:energy}.
	\end{align}
\end{subequations}
Here $\rho$ is the density, $\vect{u}=[u,v,w]^T$ is the velocity vector, $\epsilon = \rho \vect{u} \cdot \vect{u}/2 + \rho c_v T$ is the total energy, $p$ is the pressure, $\tau = \mu \left( \nabla \vect{u} + \nabla \vect{u}^T \right)  -   (2 / 3) \mu ( \nabla \cdot \vect{u} ) \text{I}$ depicts the Cauchy viscous stress tensor and $\vect{q} = - k \nabla T$ is the heat flux according to Fourier's law. The system closes with an equation of state, which for an ideal gas reads $p = (\gamma -1) (\epsilon - \rho \vect{u} \cdot \vect{u}/2 )$.

Equations $\eqref{eq:cont} - \eqref{eq:energy}$ can be written as,
\begin{equation}
	\partial_t \vect{U} + \nabla \cdot ( \mathcal{F}_c - \mathcal{F}_v ) = \vect{0},
	\label{eq:conservation_law_cNS}
\end{equation}
in terms of the conservative variables $\vect{U} = [\rho, \rho \vect{u}, \epsilon]^T$, inviscid and viscous fluxes with entries,
\begin{equation}
	\begin{aligned}
		                                                                                                                                            &
		\mathcal{F}_c = \big[\, \rho \vect{u} \left|\, \rho \vect{u} \otimes \vect{u} + p \text{I}\, \right|\, ( \epsilon + p) \vect{u} \, \big]^T, &
		\mathcal{F}_v = \big[\, \vect{0}\, \left| \, \tau\, \right|\, \tau \vect{u} - \vect{q}\, \big]^T,
		\label{fluxes}
	\end{aligned}
\end{equation}
respectively.

The conservation laws are followed by appropriate boundary conditions. These are subsonic inflow-outflow conditions on the far-field boundary $\Gamma_{\rm ff}$ and adiabatic no-slip conditions on the solid body's surface $\Gamma_{\rm w}$, with $\Gamma_{\rm ff} \cup \Gamma_{\rm w} = \partial \Omega = \Gamma_b$, where $\Omega$ spans the entire fluid domain.

\subsection{An ALE Runge-Kutta Discontinuous Galerkin Methodology}

In the following, the cNS is discretized using a Discontinuous Galerkin (DG) method in space. Also, an explicit time integration strategy is employed while taking into account mesh deformation.

To apply a DG method, firstly, $\Omega$ is discretized into a mesh $\{{\cal T}\}= \{ K_i,\ i=1,\dotsc,N_{\rm e}\}$ containing triangular elements of volume $\mathcal{D}^K$, with $\Omega_h = \cup_{K \in {\cal T}} \mathcal{D}^K$. Also $\Gamma_{\rm int} = \{ K_+\cap K_-,\ \forall K_+,K_- \in \mathcal{T}\}$ denotes all the interior faces of $\cal T$. Then, the numerical solution is assumed to be polynomial of order at most $p$ inside every element $K$, yet double-valued on $\Gamma_{\rm int}\cup \Gamma_{b}$. The element-wise solution is expressed in terms of a polynomial basis $\psi_n^K(\vect{x})$ as,
\begin{equation}
	\vect{U}^K_h(\vect{x},t) = \sum_{n = 1}^{N_p} \vect{U}^K_n(t) \ \psi_n^K (\vect{x}).
	\label{eq:series_U}
\end{equation}
In \eqref{eq:series_U}, $\psi_n^K(\vect{x})$ is arbitrary; however, in our solver nodal bases were used.

The DG discretization inside a single element $K$ is formalized by,
\begin{equation}
	\int_{\mathcal{D}^K(t)} \partial_t \vect{U}^K_h \psi_n^K   \Idiffx - \int_{\mathcal{D}^K(t)} \mathcal{F}^K_h \cdot \nabla \psi_n^K \Idiffx +
	\int_{\partial \mathcal{D}^K(t)} \vect{n}^K \cdot ( \mathcal{F}^K_h )^*  \psi_n^K  \Idiffx = \vect{0},
	\label{eq:weak_form_wo_ALE}
\end{equation}
with $\mathcal{F}^K_h$ being the total flux computed on $\vect{U}^K_h$ and $( \mathcal{F}^K_h )^* $ referring to the numerical flux that depends not only on $K$ but also on its immediate neighbors.
Since every element $K$ is able to deform during time marching, conservation can be retrieved using the general transport theorem,
\begin{equation}
	\frac{d}{dt} \int_{\mathcal{D}(t)} \vect{f} \Idiffx = \int_{\mathcal{D}(t)} \partial_t \vect{f} \Idiffx  + \int_{\partial \mathcal{D}(t)} \vect{n} \cdot (\mathbf{v} \otimes \vect{f}) \Idiffx.
	\label{eq:rtt}
\end{equation}
Now, by using \eqref{eq:rtt} alongside \eqref{eq:weak_form_wo_ALE}, the discretized form reads,
\begin{equation}
	\begin{aligned}
		 & \frac{d}{dt} \int_{\mathcal{D}^K(t)} \vect{U}^K_h \psi_n^K   \Idiffx - \int_{\mathcal{D}^K(t)} (\mathcal{F}^K_h - \mathbf{v}^K \otimes \vect{U}_h^K) \cdot \nabla \psi_n^K \Idiffx
		+ \int_{\partial \mathcal{D}^K(t)} \vect{n}^K \cdot ( \mathcal{F}^K_h - \mathbf{v}^K \otimes \vect{U}_h^K )^*  \psi_n^K  \Idiffx = \vect{0}.
	\end{aligned}
	\label{eq:ale}
\end{equation}
The numerical flux, which is included in the right integral of \eqref{eq:ale} consists of a Roe approximate Riemann solver which includes the moving mesh term $\mathbf{v}^K \otimes \vect{U}_h^K$ for the advection part, and an Interior Penalty flux for the diffusion part.
Further details on the numerical fluxes used as well as the implementation of the boundary conditions can be found in the work of Zafeiris et al. \cite{zafeiris2025parallel}, where the original solver is presented.
For the evaluation of the above integrals, a reference element $\hat{K} = \{[1,-1]^T, [-1,1]^T,[-1,-1]^T\}$ of volume $\cal D$ is used. Then, a mapping is defined, which can be expressed as a continuous diffeomorphism $\mathcal{G}^K: \mathbb{R}^2\times \mathbb{R}^+ \rightarrow \mathbb{R}^2$, that maps any point $\vect{r}=(r,s)$ of $\hat{K}$ at time $t$ to the physical coordinate $\vect{x}$ of the Eulerian frame. In this work, we use triangular elements which result in a  constant \footnote{in space and per element $K$} mapping to $\hat{K}$. This
offers a great memory advantage; see \cite{zafeiris2025parallel}.

\begin{figure}[h]
	\centering
	\includegraphics[width=0.65\textwidth,keepaspectratio]{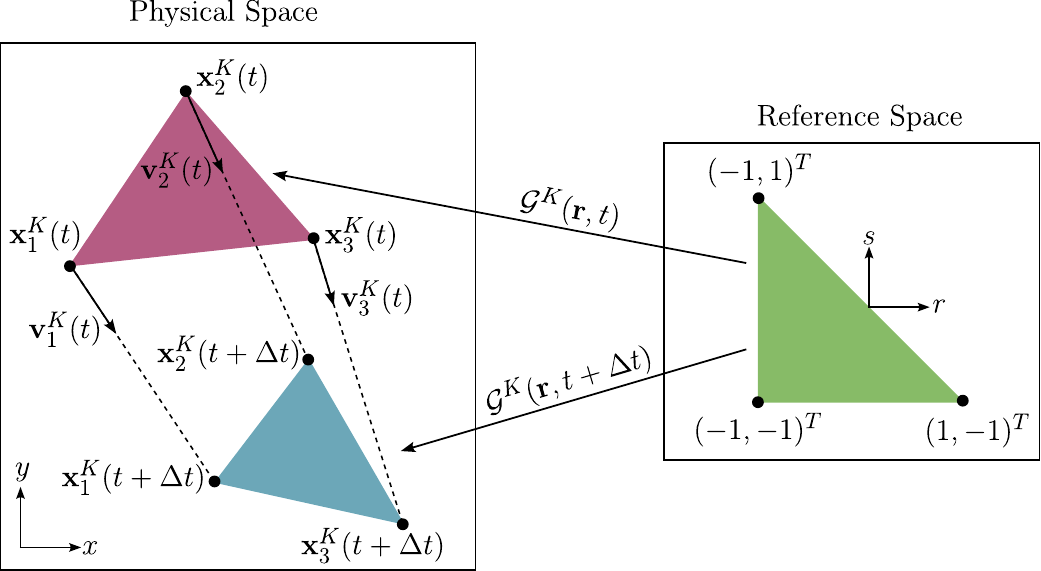}
	\caption{Schematic of an arbitrary deformation of an element $K$. The mapping from the reference space to the physical space is illustrated using the reference element $\hat{K}$. }\label{figs:triangle_transform}
\end{figure}

The mapping $\mathcal{G}^K$ is expressed in terms of the vertices of each triangle $\mathbf{x}^K_i(t), i=1,2,3$ (see Figure~\ref{figs:triangle_transform}). Since every vertex has its own arbitrary motion, $\mathcal{G}^K$ should be time-dependent. In particular,
\begin{equation}
	\mathcal{G}^K (\vect{r},t) = -\frac{r+s}{2} \mathbf{x}^K_1 (t) +\frac{1+r}{2} \mathbf{x}^K_2 (t) + \frac{1+s}{2} \mathbf{x}^K_3 (t).
	\label{eq:evaluate_jacobian}
\end{equation}
In Figure~\ref{figs:triangle_transform} we provide a schematic on how the arbitrary deformation is handled with the use of the mapping.
The polynomial basis composed with the mapping to the reference coordinates $\vect{r}$ reads,
\begin{equation}
	\psi_n^K \left( \mathcal{G}^K (\vect{r},t) \right) = \psi_n(\vect{r}).
	\label{eq:ref_base}
\end{equation}
\eqref{eq:ref_base} implies that after the use of mapping ${\cal G}^K$, the composition $\psi_n(\vect{r})$ is common for all $K \in \{\mathcal{T}\}$ and constant in time. On the other hand, the transformation's Jacobian matrix,
\begin{equation}
	\mathcal{G}^K_{\vect{r}}(t) = \nabla_{\vect{r}} \mathcal{G}^K (\vect{r} , t), \label{eq:jac_def_DG}
\end{equation}
is time dependent. Rewriting now \eqref{eq:ale} using $\mathcal{G}^K_{\vect{r}}(t)$ and $\mathcal{J}^K(t)$ as the determinant of $\mathcal{G}^K_{\vect{r}}(t)$, which is hereinafter referred to as the Jacobian for brevity, we have,
\begin{equation}
	\begin{aligned}
		 & \frac{d}{dt} \int_{\mathcal{D}} \vect{U}^K_h \psi_n  \mathcal{J}^K (t) \Idiffr
		- \int_{\mathcal{D}} ( \mathcal{F}^K_h - \mathbf{v}^K \otimes \vect{U}_h^K ) \cdot {\left( \mathcal{G}^K_{\vect{r}} (t) \right)}^{-T} \nabla_{\vect{r}} \psi_n \mathcal{J}^K (t) \Idiffr \\
		 & \quad + \int_{\partial \mathcal{D}} \vect{n}^K \cdot ( \mathcal{F}^K_h - \mathbf{v}^K \otimes \vect{U}_h^K )^*  \psi_n \mathcal{J}^K_s (t) \Idiffr  = 0,
		\label{eq:weak_form_with_ALE}
	\end{aligned}
\end{equation}
whereby the physical gradient operator is transformed using the gradient operator in the reference space, $\nabla \equiv \left( \mathcal{G}^K_{\vect{r}} \right)^{-T} \nabla_{\vect{r}}$.
The nodal base as well as the nodal interpolation points for the reference element $\hat{K}$ are chosen based on \cite{hesthaven2008nodal}.

Let $\vect{\mathcal{U}}$ contain the total DG degrees of freedom $\forall K \in \{\mathcal{T}\}$. For the triangular elements used here, the mass matrix per element is evaluated as $\mathcal{M}^K (t) = \mathcal{J}^K(t) \mathcal{M}$, with $\mathcal{M}_{nm} = \int_{\mathcal{D}} \psi_n \psi_m \Idiffr$. Thus, only the Jacobian needs to be stored for all elements with the mass matrix $\mathcal{M}$ only needed to be computed once and stored for \textit{all} elements.
For all elements $K \in {\cal T}$, the product $\vect{\mathcal{J}}(t) \vect{\mathcal{M}}$ is the block-diagonal mass matrix. Here the diagonal matrix $\vect{\mathcal{J}}(t)$ contains the Jacobians' determinant while $\vect{\mathcal{M}}$ assembles $N_e$-times the same matrix $\cal M$.
Therefore $\vect{\cal M}^{-1}$ is readily available. The semi-discrete representation reads,
\begin{equation}
	\frac{d}{dt} (\vect{\mathcal{J}}(t) \vect{\mathcal{U}}) = \vect{\mathcal{M}}^{-1}
	\vect{\mathcal{R}} (\vect{x}_v(t), \vect{\mathcal{U}}),
	\label{eq:DG_disc_semi}
\end{equation}
whereby $\vect{x}_v(t)$ are the  coordinates of the nodes comprising the mesh.

\subsubsection{Free-Stream Preservation and the Geometric Conservation Law}\label{sc:fsp_gcl}

In ALE formulations, it is crucial to approximate a constant solution robustly and preferably up to  machine precision. To enforce this criterion, we demand that \eqref{eq:weak_form_with_ALE} equals zero after inserting a constant $\bar{\vect{U}}$. Then, we have:
\begin{align}
	\bar{\vect{U}} \Big(
	\frac{d}{dt} \int_{\mathcal{D}} \psi_n \, \mathcal{J}^K \, \Idiffr
	 & + \int_{\mathcal{D}} \mathbf{v}^K \cdot \left( \mathcal{G}^K_{\vect{r}} \right)^{-T}
	\nabla_{\vect{r}} \psi_n \, \mathcal{J}^K \, \Idiffr
	- \int_{\partial \mathcal{D}} \vect{n}^K \cdot \mathbf{v}^K \, \psi_n \, \mathcal{J}^K_s \, \Idiffr
	\Big) \nonumber                                                                         \\
	 & - \mathcal{F}(\bar{\vect{U}}) \cdot \Big(
	\int_{\mathcal{D}} \left( \mathcal{G}^K_{\vect{r}} \right)^{-T}
	\nabla_{\vect{r}} \psi_n \, \mathcal{J}^K \, \Idiffr
	- \int_{\partial \mathcal{D}} \vect{n}^K \, \psi_n \, \mathcal{J}^K_s \, \Idiffr
	\Big) = \vect{0}. \label{eq:freestream}
\end{align}
Concerning \eqref{eq:freestream}, the expression multiplied with $\mathcal{F}(\bar{\vect{U}})$ is trivially zero after applying the divergence theorem. The remaining expression, after a few algebraic manipulations, see~\ref{sec:appendix-gcl}, is an \textit{Ordinary Differential Equation} (ODE) which reads,
\begin{equation}
	\frac{d}{dt} \mathcal{J}^K - C^K (t) \mathcal{J}^K = 0, \ C^K (t) = \nabla_{\vect{r}} \cdot \big( \big( \mathcal{G}^K_{\vect{r}} (t) \big)^{-1} \mathbf{v}^K (t) \big). \label{eq:gcl}
\end{equation}
\eqref{eq:gcl} is called the Geometric Conservation Law and, due to the fact that the mapping $\mathcal{G}^K_{\vect{r}}$ is affine, i.e., for de facto triangular elements, it is trivially satisfied; see~\ref{sec:appendix-gcl}. However, numerical time integration of \eqref{eq:DG_disc_semi} renders \eqref{eq:gcl} non-exact and therefore the discrete GCL does not hold. To mitigate this, Persson et al. \cite{persson2009discontinuous} suggested that, $\mathcal{J}^K$ could also be numerically marched in time along with the main system.


In contrast to $\mathcal{J}^K \big( t^{(n+1)} \big)$, which is explicitly evaluated through \eqref{eq:evaluate_jacobian} at $t = t^{(n+1)}$, we use $\overline{\mathcal{J}^K}$ to represent the numerical approximation of $ \mathcal{J}^K \big( t^{(n+1)} \big)$. To that end, \eqref{eq:gcl} can be written in the following way:
\begin{equation}
	\frac{d}{dt} \overline{\mathcal{J}^K} - C^K (t) \mathcal{J}^K = 0. \label{eq:GCL_eq}
\end{equation}
\eqref{eq:GCL_eq} is discretized using the same  time integration scheme as the main system. By employing this strategy, conservation of the free stream is satisfied up to the machine precision.
$\overline{\mathcal{J}^K}$ then replaces $\mathcal{J}^K$ in the first term of \eqref{eq:weak_form_with_ALE} to yield,
\begin{equation}
	\begin{aligned}
		\frac{d}{dt} \int_{\mathcal{D}} \vect{U}^K_h \psi_n  \overline{\mathcal{J}^K} \Idiffr & - \int_{\mathcal{D}} ( \mathcal{F}^K_h - \mathbf{v}^K \otimes \vect{U}_h^K ) \cdot {\left( \mathcal{G}^K_{\vect{r}} \right)}^{-T} \nabla_{\vect{r}} \psi_n \mathcal{J}^K \Idiffr \\
		                                                                                      & + \int_{\partial \mathcal{D}} \vect{n}^K \cdot ( \mathcal{F}^K_h - \mathbf{v}^K \otimes \vect{U}_h^K )^*  \psi_n \mathcal{J}^K_s \Idiffr = 0.
	\end{aligned}
	\label{eq:weak_form_with_ALE_GCL}
\end{equation}

In summary, the total semi-discrete ALE DG system with the GCL reads,
\begin{equation}
	\begin{aligned}
		 & \frac{d}{dt}
		\bigg\{ \begin{array}{c}
			        \overline{\vect{\mathcal{J}}} \\
			        \overline{\vect{\mathcal{J}}}
			        \vect{\mathcal{U}}
		        \end{array}\bigg\}
		=
		\bigg\{ \begin{array}{c}
			        \vect{C} \vect{J} \\
			        \vect{\mathcal{M}}^{-1} \vect{\mathcal{R}} \left( \vect{x}_v, \vect{\mathcal{U}} \right)
		        \end{array} \bigg\},
	\end{aligned}
	\label{eq:disc_DG_w_GCL}
\end{equation}
whereby $\vect{C}$ assembles $C^K$ for all $K \in \{\mathcal{T}\}$.

\subsubsection{RBF Grid Deformation}
For the grid deformation, the Radial Basis Function method is adopted. The RBF grid deformation technique has been applied in a wide range of applications \cite{zhong2020efficient,de2007mesh,xie2017efficient}. Here, a brief overview is given following the work of Rendall and Allen \cite{rendall2009efficient}. The chosen basis function, $\phi(\xi),\ \xi \in [0,1]$, is Wendland's $C2$ and all distances are measured using a weighted Euclidean norm,
\begin{equation}
	\| \Delta \vect{x}  \|_{\vect{w}} := \sqrt{ (w_x \Delta x)^2 + (w_y \Delta y)^2 + (w_z \Delta z)^2 }, \quad \vect{w} = [w_x,w_y,w_z]^T,
	\label{Weighted_Eucl_Norm}
\end{equation}
with $\vect{w}$ being free parameters. For symmetric problems, all weights are typically set to one. However, when one spatial direction undergoes very large deformation, relative to the other, its corresponding weight can be chosen smaller than one in order to relax and stabilize the deformation.

The deformation is numerically applied through the  matrix $H$ defined as $H = A M^{-1}$. The interpolation matrix $M$ and calculation matrix $A$ are defined as
\begin{align}
	 & M_{ij} = \phi\left( {\| \vect{x}_{s_i} - \vect{x}_{s_j} \|}_{\vect{w}} / R\right), & A_{kj} = \phi\left( {\| \vect{x}_{v_k} - \vect{x}_{s_j} \|}_{\vect{w}} / R\right),
\end{align}
whereby the index $s_i$ refers to the $i$-th wall boundary node, the index $v_k$ refers to the $k$-th grid node and $R$ is the support radius, which is a free parameter.

The calculation matrix $M$ is symmetric and positive definite (see \cite{rendall2009efficient}), which guarantees numerical stability and allows its inversion to be directly performed using methods such as the Cholesky decomposition. In the present work, the number of wall-boundary nodes does not exceed 500; therefore, neither memory limitations nor numerical conditioning issues are encountered. When memory requirements become significant, greedy algorithms \cite{rendall2010reduced} can be employed to reduce the amount of stored data.

\boldparagraph{Remark}
Since the solver is implemented in parallel on domain-decomposed blocks through MPI, mesh deformation of the entire domain is not a straightforward process. Specifically, $M$ refers to all the wall boundary nodes of the domain and its direct storage is inefficient.
However, the column number of $A$ is the number of grid nodes per block, which decreases as the number of blocks increases. Hence, $H = A M^{-1}$ should be performed in a pre-processing step for every single block and only $H$ is eventually stored.

The volume displacements of the grid nodes $\Delta \vect{x}_v$ are,
\begin{equation}
	\Delta \vect{x}_v = H \Delta \vect{x}_s,
\end{equation}
with $\Delta \vect{x}_s$ being the surface nodes' displacements. Their evaluation is described in \ref{subsub:timediscrete}.
\subsubsection{Rigid Body Dynamics} \label{sc:RBD}
To conclude the fluid-structure coupling, the motion of each rigid body must be determined from the aerodynamic loads. The \textit{Rigid Body Dynamics} (RBD) are described by a system of second-order ODEs for all solid bodies, as follows,
\begin{equation}
	M_b \ddot{\vect{x}}_b (t) + C_b \dot{\vect{x}}_b (t) + K_b \vect{x}_b (t) = \vect{F}_f \left(  \vect{x}_s(t) , \vect{\mathcal{U}} (t)  \right). \label{eq:rbd_eq}
\end{equation}
If $N_{\rm bd}$ is the total number of DoF of oscillation, then $\vect{x}_b = [x_{b,1}, \dotsc,  x_{b,N_{\rm bd}}]$ is defined as the solution vector containing $x,y$ displacements for all bodies; the displacements are measured from the bodies' initial position. Similarly, $\dot{\vect{x}}_b$ are the velocities; also a solution of the second-order system of ODEs.

Additionally, $M_b, C_b$ and $K_b$ are the mass, damping and stiffness matrices respectively. If no dynamic inter-body coupling is present, then all three matrices are block-diagonal. For $M_b$, its diagonal entries contain the mass of each body while entries of $C_b$ and $K_b$ are damping and spring constants for each DoF per body with respect to $\vect{x}_b$. Lastly, $\vect{F}_f$ is the vector containing the aerodynamic forces. Let $\vect{F}_{f,i}$ refer to the loads of the $i$-th body, then,
\begin{equation}
	\vect{F}_{f,i} = \int_{S_i} \big( p \mathrm{I} + \tau \big) \vect{n} \Idiffx,
	\label{eq:force_int}
\end{equation}
with $S_i = S_i (\vect{x}_s)$ being the surface of the $i$-th body.

\eqref{eq:rbd_eq} is numerically solved using a Newmark-$\beta$ method with $\beta=0.25$ and $\gamma = 0.5$ coefficients. If $i$ corresponds to the surface node index of the $k$-th body, then $\Delta \vect{x}_{s_i} = \vect{x}_{b,k}$, whereby $\vect{x}_{b,k}$ corresponds to the degrees of freedom of the $k$-th body. This can be described mathematically via the matrix $L$ of size $N_s \times N_b$ with $\Delta \vect{x}_{s} = L \vect{x}_{b} $.

\subsubsection{Time Discretization and Assembly}\label{subsub:timediscrete}

The time-marching of \eqref{eq:disc_DG_w_GCL} is performed using an explicit low-storage Strong Stability-Preserving Runge-Kutta (SSPRK) scheme. In particular, a five-stage method is adopted as in  \cite{niegemann2012efficient}. The implementation of the SSPRK discretization of \eqref{eq:disc_DG_w_GCL} for calculating the solution at $t^{(n+1)} = t^{(n)} + \Delta t$ is described by Algorithm~\ref{al:ssperk}.

For notational clarity, any quantity $A$ evaluated at time $t^{(n)}$ is denoted by $A^{(n)}$, and analogously for $t^{(n+1)}$. The value of $A$ at stage $i$ is written as $A_i$, indicating that all of its inputs correspond to the solution state at that Runge-Kutta stage.

\begin{algorithm}[h]
	\caption{Low-Storage SSPRK Time Integration and Coupling with Deforming Grid and RBD}
	\setstretch{1.2}
	\begin{algorithmic}[1]
		\State
		$\vect{\mathcal{U}}_0 \leftarrow \vect{\mathcal{U}}^{(n)}$;\quad
		$\vect{\mathcal{K}}_0 \leftarrow \vect{\mathcal{U}}^{(n)}$;\quad
		$\overline{\vect{\mathcal{J}}}_0 \leftarrow \overline{\vect{\mathcal{J}}}^{(n)}$;\quad
		$\vect{\mathcal{N}}_0 \leftarrow \overline{\vect{\mathcal{J}}}_0$;\quad
		$\vect{x}_{v,0} \leftarrow \vect{x}_{v}^{(n)}$;\quad
		$\vect{\mathcal{P}}_0 \leftarrow \vect{x}_{v,0}$  \Comment{Initialization}
		\State $\vect{F}_f^{(n)} = \mathtt{CALCULATE\_FORCES} (  \vect{x}_s^{(n)} , \vect{\mathcal{U}}^{(n)} )$
		\State $\dot{\vect{x}}_b^{(n+1)} = \mathtt{RBD\_SOLVER} (\mathbf{F}_f^{(n)})$
		\State $\Delta \vect{x}_s \leftarrow L ( \vect{x}_b^{(n)} + \Delta t \dot{\vect{x}}_b^{(n+1)} ) $
		\State $\Delta \vect{x}_v^{(n+1)} = H \Delta \vect{x}_s$
		\State $\mathbf{v}^{(n+1)} = ( \Delta \vect{x}_{v}^{(n+1)}  - \Delta \vect{x}_{v}^{(n)} ) / \Delta t$
		\For{$i = 1, \dots, N_s$}
		\State $\vect{\cal R}_{i} \leftarrow
			\vect{\mathcal{R}}( \vect{x}_{v,i-1} ,  \vect{\mathcal{U}}_{i-1}) $
		\State $\vect{\mathcal{P}}_{i} \leftarrow \alpha_i \, \vect{\mathcal{P}}_{i-1} + \Delta t \, \mathbf{v}^{(n+1)}$
		\State $\vect{x}_{v,i} \leftarrow \vect{x}_{v,i-1} + \beta_i \, \vect{\mathcal{P}}_i$
		\Comment{Position update}
		\State $\vect{\mathcal{N}}_i \leftarrow \alpha_i \, \vect{\mathcal{N}}_{i-1} + \Delta t \, \vect{C}_{i-1} \, \overline{\vect{\mathcal{J}}}_{i-1}$
		\State $\overline{\vect{\mathcal{J}}}_i \leftarrow \overline{\vect{\mathcal{J}}}_{i-1} + \beta_i \, \vect{\mathcal{N}}_i$
		\Comment{Time-marching of the numerical Jacobian}
		\State $\vect{\mathcal{J}}_i \leftarrow \overline{\vect{\mathcal{J}}}_i$\ ;\ $\vect{\mathcal{J}}_{i-1} \leftarrow \overline{\vect{\mathcal{J}}}_{i-1}$
		\State $\vect{\mathcal{K}}_i \leftarrow \alpha_i \,
			(\vect{\cal J}^{-1}_i \vect{\cal J}_{i-1})
			\, \vect{\mathcal{K}}_{i-1}
			+ \Delta t \, \vect{\cal J}^{-1}_{i-1}
			\vect{\mathcal{M}}^{-1}
			\vect{\cal R}_{i}$
		\State $\vect{\mathcal{U}}_i \leftarrow
			(\vect{\cal J}^{-1}_i \vect{\cal J}_{i-1})
			\vect{\mathcal{U}}_{i-1} + \beta_i \, \vect{\mathcal{K}}_i$
		\Comment{Solution update}
		\EndFor

		\State $\vect{\mathcal{U}}^{(n+1)} \leftarrow \vect{\mathcal{U}}_{N_s}$
	\end{algorithmic} \label{al:ssperk}
\end{algorithm}

The quantities $\vect{\mathcal{K}}_i$, $\vect{\mathcal{N}}_i$, and $\vect{\mathcal{P}}_i$, correspond to the solution vector $\vect{\mathcal{U}}$, the Jacobian vector $\vect{\mathcal{J}}$ and the volume node vector $\vect{x}_v$, respectively. They depict the intermediate values of those variables at each Runge-Kutta stage $i$. The scheme is low-storage, since only two stages are stored regardless of the RK scheme's total stage number.

The boundary velocity $\dot{\vect{x}}_b^{(n+1)}$ at time $t^{(n+1)}$ is used to compute the surface nodes' displacements  as seen in the right hand side of line 4 of Algorithm~\ref{al:ssperk}.

The coupling with the rigid-body dynamics (RBD) solver is introduced through the grid velocity $\mathbf{v}^{(n+1)}$ using a first-order finite difference approximation. The grid is deformed consistently with the numerical scheme, as shown in lines 6-8 of Algorithm~\eqref{al:ssperk}, by assuming the grid velocity to remain constant across all RK stages.

It has to be mentioned that in a non-ALE DG-discretized system, all DG matrices are evaluated only at the beginning of the program (this could be a pre-processing level) and they are reused at every time step. However, when applying mesh deformation, they have to be re-evaluated. An advantage of triangular meshes is that the total amount of extra evaluations due to ALE involve only the calculation of the Jacobian matrix for every element, as per \eqref{eq:evaluate_jacobian}.

%% file: numerical_results.tex
To  evaluate the presented ALE-DG framework, this section presents a series of progressively complex numerical investigations.  First, we verify the \textit{Free Stream Preservation} (see \cite{persson2009discontinuous} and \cite{Nguyen2010}) (FSP) property of the GCL-based ALE nodal IP RKDG framework, which quantifies, in other words, the ability of the method to approximate a free-stream solution over a deformable mesh compared to the machine precision. The solver is then applied to complex multi-body VIV configurations: first, a two-cylinder tandem arrangement undergoing cross-flow oscillations, and second, a three-cylinder tandem system with $x-y$ translational freedom. To confirm the accuracy and low-dissipative nature of the ALE-DG framework, the resulting coupled dynamics and wake structures are evaluated against results available in the literature. All the cases considered  are in the incompressible regime, therefore, we use a nominal Mach number $\mathrm{Ma} = U / \sqrt{ \gamma p_{\infty} / \rho_{\infty}}=0.15$, to suppress compressibility effects.

\subsection{Numerical Validation of the Free Stream Preservation}\label{GCL_validation}

\begin{figure}[h]
	\centering
	\begin{subfigure}[t]{0.3\textwidth}
		\centering
		\includegraphics[width=\textwidth,keepaspectratio]{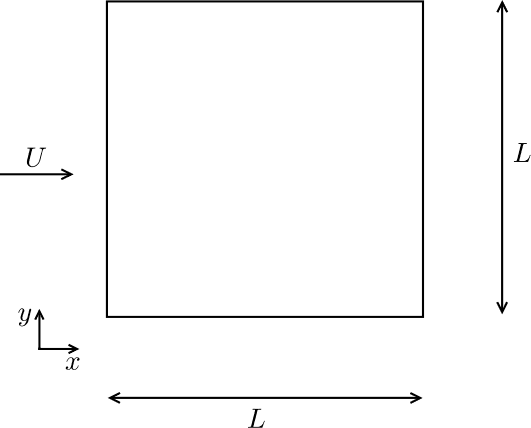}
		\caption{}
		\label{subfig:GCL_problem}
	\end{subfigure}
	\hspace{0.05\textwidth}  
	\begin{subfigure}[t]{0.55\textwidth}
		\centering
		\includegraphics[width=\textwidth,keepaspectratio]{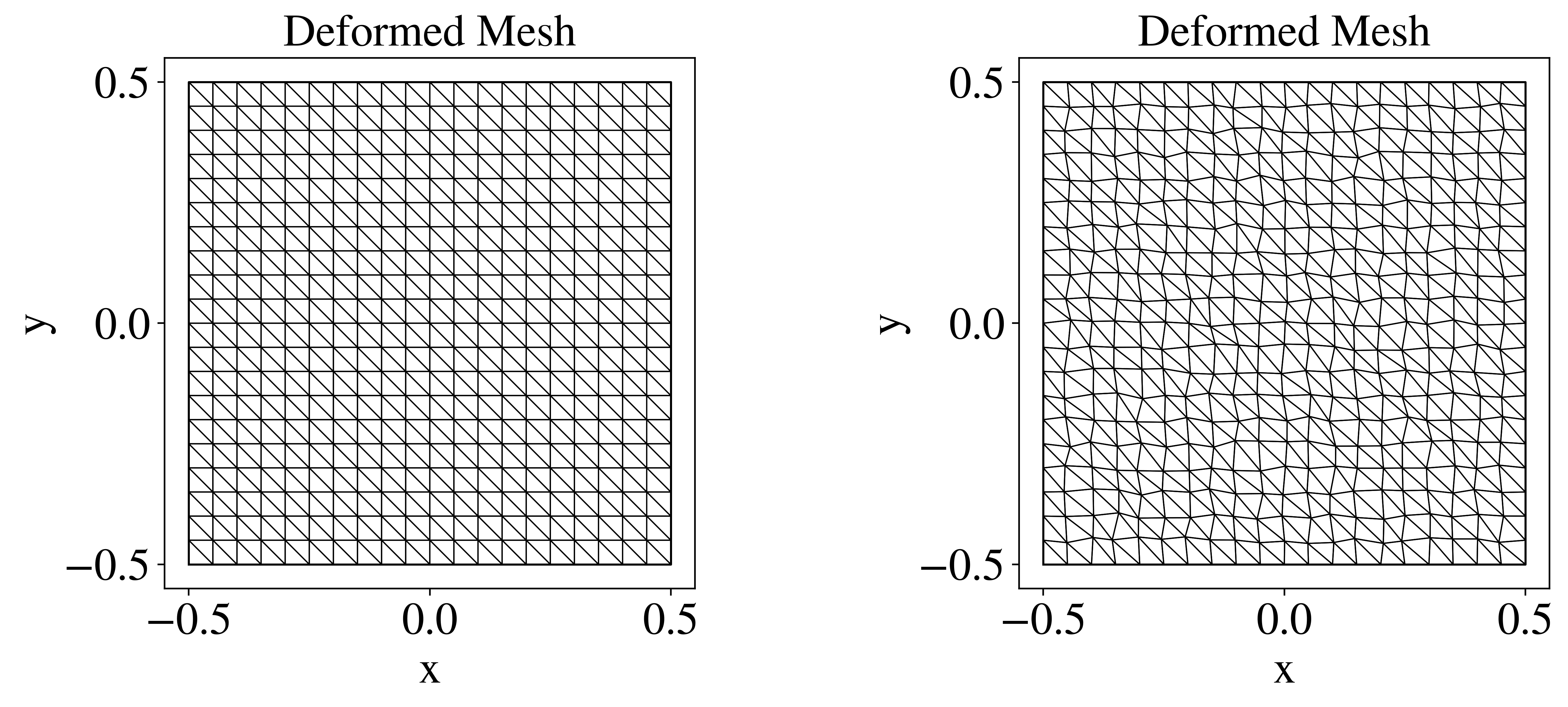}
		\caption{}
		\label{subfig:GCL_grid}
	\end{subfigure}
	\caption{Free Stream Preservation problem and corresponding computational grid. On the left, the computational setup is defined. On the right the mesh is illustrated in both the original and an instantaneous deformed state.}
	\label{fig:GCL_combined}
\end{figure}

To explore the effect of the GCL on the FSP, the following setup is considered. A rectangular grid of size $L \times L$ is used. On each side of the rectangle an inflow - outflow condition is applied. The computational setup  can be seen in Figure \ref{subfig:GCL_problem}. For this simulation $L$ and $U$ were both chosen to be equal with 1. The grid consists of 800 triangular elements, each one initially being a right isosceles triangle as seen in the left image Figure \ref{subfig:GCL_grid}. Also, the internal nodes of the mesh are able to move in a random fashion, with a displacement following a uniform distribution, as seen in the right image of Figure \ref{subfig:GCL_grid}, boundary nodes are clamped.

The GCL requires that an initially uniform solution remains constant throughout the time marching and grid deformation process. The accuracy of this free-stream preservation is evaluated for any flow variable $v$ via the $\mathrm{minmax}(\cdot)$ function, given by,
\begin{equation}
	\mathrm{minmax}(v) := \max_{t \in I} \big( \max_{\vect{x} \in \Omega } v(\vect{x},t) - \min_{\vect{x} \in \Omega} v(\vect{x},t) \big),
\end{equation}
with $I=[0,t_f)$ the time interval in which the simulation is performed. For this simulation, the fluid's density $\rho$ will be used to measure the norm. In order to ensure that the GCL holds various polynomial orders, we considered $p=1,3,5$.
\begin{figure}[h]
	\centering
	\includegraphics[width=0.50\textwidth,keepaspectratio]{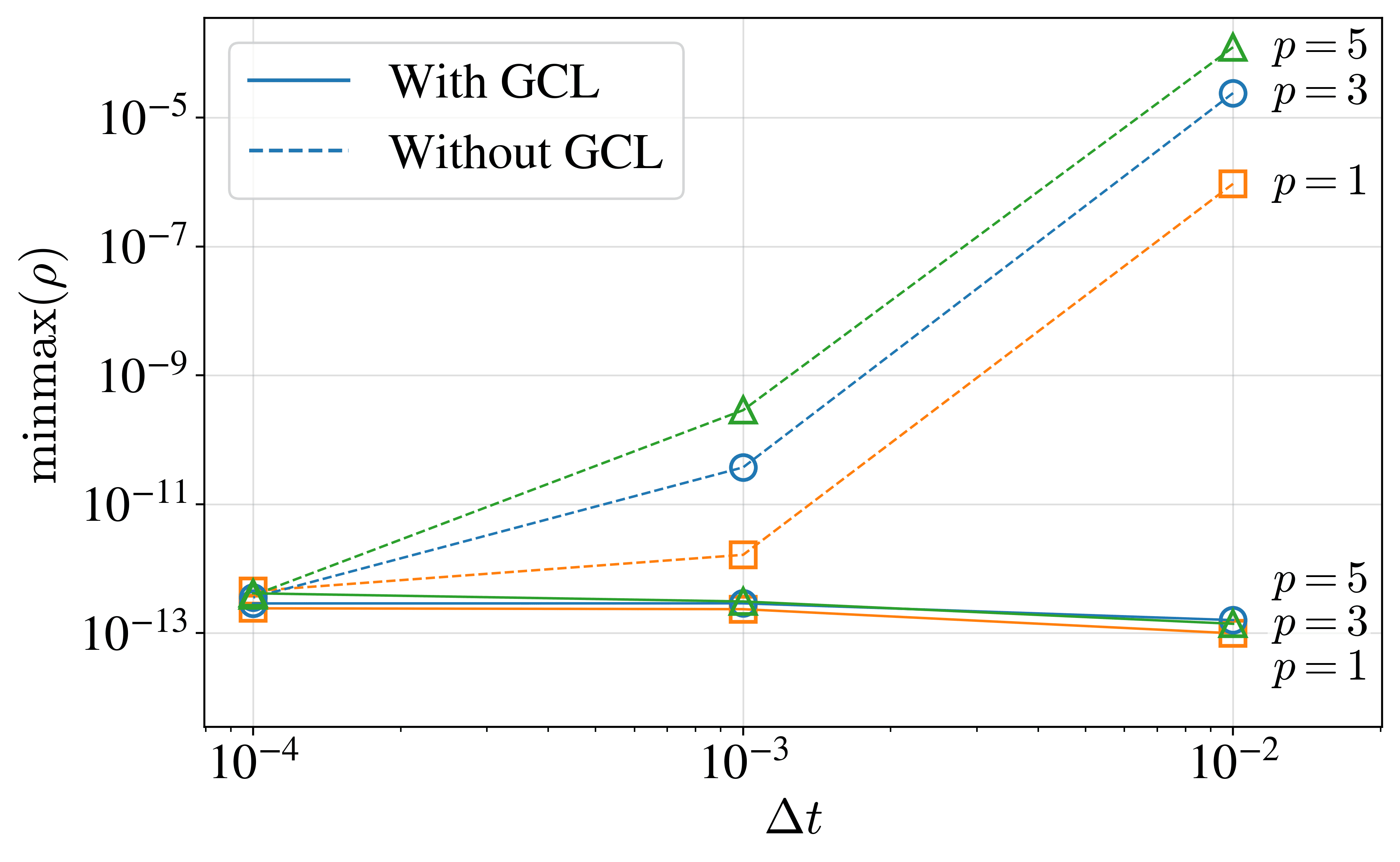}
	\caption{Results of the Free Stream Preservation simulations. $\triangle$ (green line) refers to $p=5$, $\bigcirc$ (blue line) to $p=3$ and $\square$ (orange line) to $p=1$.}\label{GCL_results}
\end{figure}

The results are presented in Figure \ref{GCL_results}. For small time steps $\Delta t$ the error induced by the grid deformation is small regardless of whether the GCL is applied or not. In the absence of the GCL treatment, increasing the time step size causes the numerical error to grow rapidly. This happens due to errors in the time integration scheme itself as discussed in detail in \ref{sc:fsp_gcl}. Conversely, when applying the GCL, we can make two observations. Firstly, numerical perturbations remain closer to machine precision overall, which is $\mathcal{O}(10^{-16})$ for our machine. Also, the error is non-increasing with respect to time-step, indicating an insensitivity to the time-step, at least for this range. Having confirmed the free-stream preservation properties of the solver, we proceed to apply the framework to coupled fluid-structure interaction problems.

\subsection{Two Cylinders in Tandem Arrangement}
The solver is now evaluated for a two-cylinder tandem arrangement restricted to a single degree of freedom (1-DoF) per cylinder in the cross-flow direction. The two cylinders are placed in close proximity and thus this configuration represents a challenging scenario to also investigate the performance of the numerical framework under large grid deformations. Our results will be compared to those available in the literature from Griffith et al. \cite{griffith2017flow} and Papadakis et al. \cite{papadakis2022hybrid}.

\begin{figure}[h]
	\centering
	\includegraphics[height=0.15\textwidth,keepaspectratio]{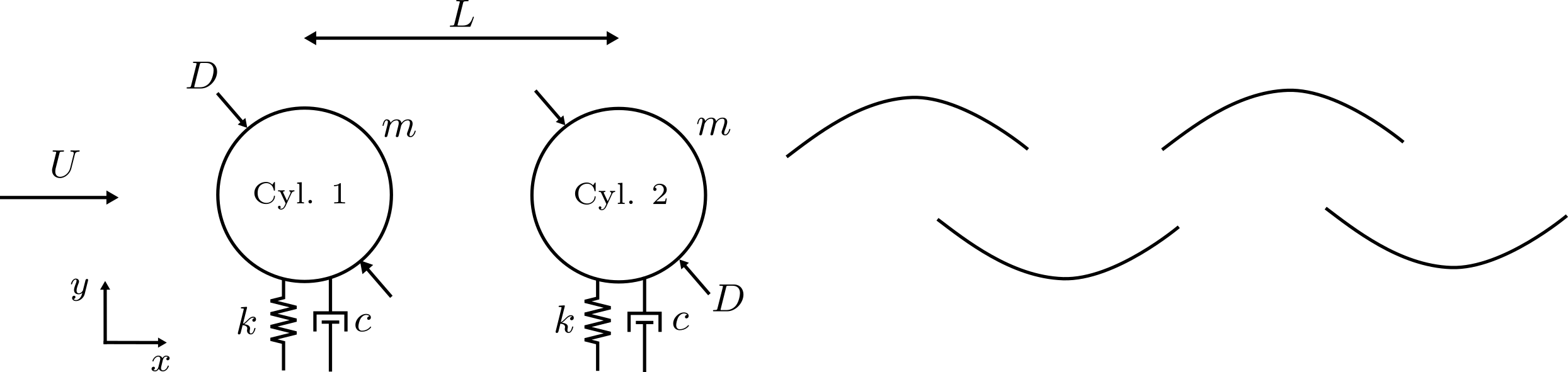}
	\caption{The two cylinders in tandem arrangement with 1 DoF in the cross-flow direction.}\label{tan_cyl_config}
\end{figure}

The tandem configuration consists of two elastically-mounted cylinders aligned parallel to a uniform flow with a fixed center-to-center spacing of $L/D = 1.5$.
Initial and boundary conditions translate into $\rm Re=200$ and $\rm Ma=0.15$ Reynolds and Mach numbers, respectively. The dynamics are governed by the mass ratio $m^* = m/m_f$, where $m_f = \rho_f \pi D^2 / 4$ represents the displaced fluid mass, alongside the spring stiffness $k$ and damping constant $c$. The system's response is primarily characterized by the reduced velocity, $U^* = U/(f_n D)$, defined using the natural frequency $f_n = \sqrt{k/m}/(2\pi)$. The reduced displacement is defined to be $y^* = y / D$. In accordance with the benchmarks of Griffith et al. \cite{griffith2017flow} and Papadakis et al. \cite{papadakis2022hybrid}, the damping coefficient is set to $\zeta = c/(4\pi m f_n) = 0$.
These parameters are integrated into rigid body dynamics equations \eqref{eq:rbd_eq}.
The arrangement can be seen in Figure \ref{tan_cyl_config}.


A sensitivity study using $p$-refinement is conducted specifically at  the reduced velocity $U^*=7$  which, as we shall see, leads to high-amplitude oscillations and pronounced fluid-structure interaction due to the low structural stiffness.

\begin{figure}[h]
	\centering
	\begin{subfigure}{0.55\textwidth}
		\centering
		\includegraphics[width=\textwidth, keepaspectratio]{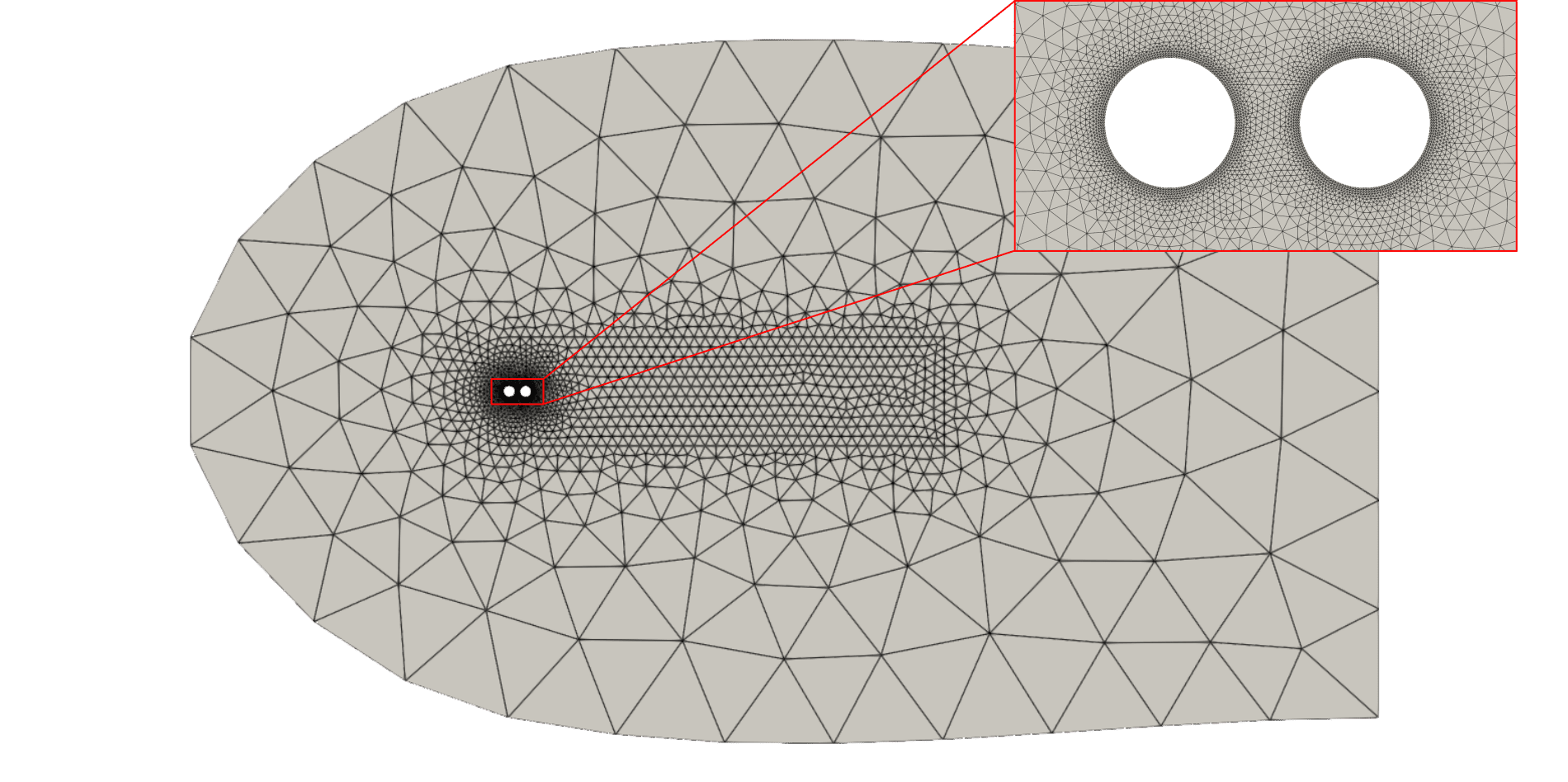}
		\subcaption{Computational grid for the tandem cylinder configuration.}
		\label{fig:tan_cyl_grid}
	\end{subfigure}
	\hfill
	\begin{subfigure}{0.42\textwidth}
		\centering
		\begin{subfigure}{\textwidth}
			\centering
			\includegraphics[width=\textwidth, keepaspectratio]{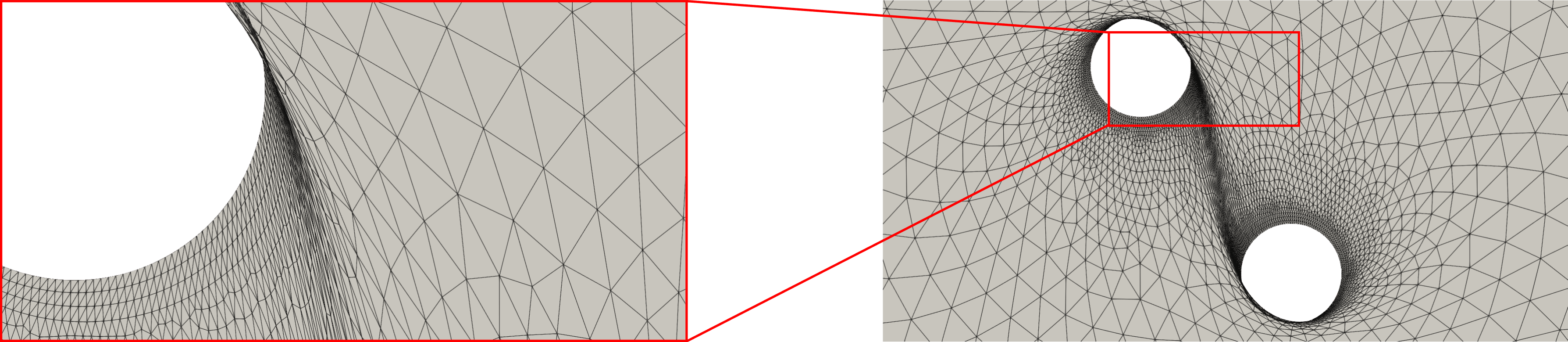}
			\subcaption{$w_y = 1.0$}
			\label{fig:deform_wy_1}
		\end{subfigure}
		\vspace{0.5em}
		\begin{subfigure}{\textwidth}
			\centering
			\includegraphics[width=\textwidth, keepaspectratio]{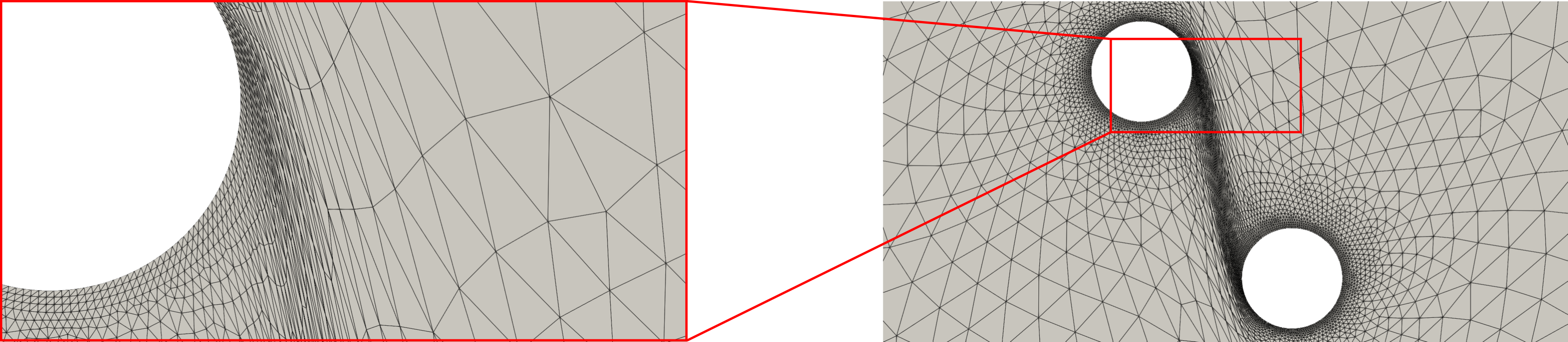}
			\subcaption{$w_y = 0.4$}
			\label{fig:deform_wy_0p4}
		\end{subfigure}
	\end{subfigure}
	\caption{The initial computational mesh (left) and the deformed mesh for large grid deformations for different values of $w_y$ (right).}
	\label{figs:deform_2_cyl}
\end{figure}

The computational domain is discretized using approximately $8,000$ elements, with $160$ nodes distributed along each cylinder's surface (see Figure \ref{fig:tan_cyl_grid}). For this specific case, the maximum amplitude of the response ($A^* \sim L/D$) poses a significant challenge for grid deformation. As it can be shown in Figure \ref{figs:deform_2_cyl}, the standard RBF method with uniform weights leads to negative cell volumes and mesh tangling near the moving boundaries (Figure \ref{fig:deform_wy_1}). To preserve mesh integrity under these conditions, the lateral deformation weights are relaxed to $w_y = 0.4$. This modification ensures a robust grid even at peak displacement (Figure \ref{fig:deform_wy_0p4}), although the resulting element skewness still imposes a restrictively small time-step imposed by the CFL condition, as shown in Table \ref{tab:cfl_dt}. There, it is also obvious that increasing the polynomial order leads to a significant reduction in the maximum allowable time step $\Delta t$, highlighting the computational overhead associated with high-order DG discretization.

After verifying that the method is indeed insensitive, subsequent simulations are performed across the range of $U^* = 4-8$.

\boldparagraph{Sensitivity Study}
The influence of the polynomial order $p$ on the solution's accuracy and computational cost is evaluated for $p=1, 3, \text{ and } 5$.
\begin{table}[h]
	\centering
	\begin{tabular}{c c c c c}
		\toprule
		$p$ & $\Delta t_{\rm min}\ [{\rm ms}]$ & $\Delta t_{\rm max}\ [{\rm ms}]$ & $\Delta t_{\rm mean} [{\rm ms}]$ \\
		\midrule
		$1$ & $0.11057$                        & $0.39579$                        & $0.19287$                        \\
		$3$ & $0.01864$                        & $0.08931$                        & $0.03534$                        \\
		$5$ & $0.00908$                        & $0.04396$                        & $0.01722$                        \\
		\bottomrule
	\end{tabular}
	\caption{CFL and time steps for each order.}
	\label{tab:cfl_dt}
\end{table}
The convergence of global physical quantities is summarized in Table \ref{tab:2_cyl_sens}. While the $p=1$ case fails to accurately capture the peak lift coefficients and dominant frequencies, the results for $p=3$ and $p=5$ show excellent agreement, with negligible differences in reduced amplitudes ($A^*_i$) and primary frequency. This convergence in the frequency domain, is further evidenced by the Normalized Power Spectra in Figure \ref{figs:insens_plot}. In particular, the primary frequency in the second-to-last column of Table \ref{tab:2_cyl_sens} for $p=1$ is far from values of $p=3$ and $p=5$. This is explained on the Normalized Power Spectra in Figure \ref{figs:insens_plot}, by observing a double-peak for $p=1$ on the left sub-figure; this is of course a product of poor accuracy.
\begin{table}[h!]
	\centering
	\begin{tabular}{|ccccc|ccc|ccc|}
		\toprule
		$p$ & $A^*_1$ & $A^*_2$ & $\max C_{L,1}$ & $\max C_{L,2}$ & \multicolumn{3}{c|}{$f_{\rm prim}^{[1]}\ {\rm [Hz]}$} & \multicolumn{3}{c|}{$f_{\rm prim}^{[2]}\ {\rm [Hz]}$}                                         \\
		\midrule
		$1$ & $0.793$ & $1.056$ & $0.559$        & $0.814$        & $0.138$                                               & $0.276$                                               & $0.407$ & $0.138$ & $0.407$ & $0.421$ \\
		$3$ & $0.778$ & $1.065$ & $0.600$        & $0.921$        & $0.138$                                               & $0.276$                                               & $0.414$ & $0.138$ & $0.283$ & $0.414$ \\
		$5$ & $0.774$ & $1.064$ & $0.594$        & $0.917$        & $0.138$                                               & $0.276$                                               & $0.414$ & $0.138$ & $0.283$ & $0.414$ \\
		\bottomrule
	\end{tabular}
	\caption{Reduced Amplitudes $A_i^*$, maximum lift coefficients $C_{L,i}^{\max}$ and top three dominant frequencies for each polynomial order $p$.}\label{tab:2_cyl_sens}
\end{table}
\newlength{\figsep}
\setlength{\figsep}{0.05\textwidth} 

\begin{figure}[h]
	\centering
	\includegraphics[width=0.40\textwidth]{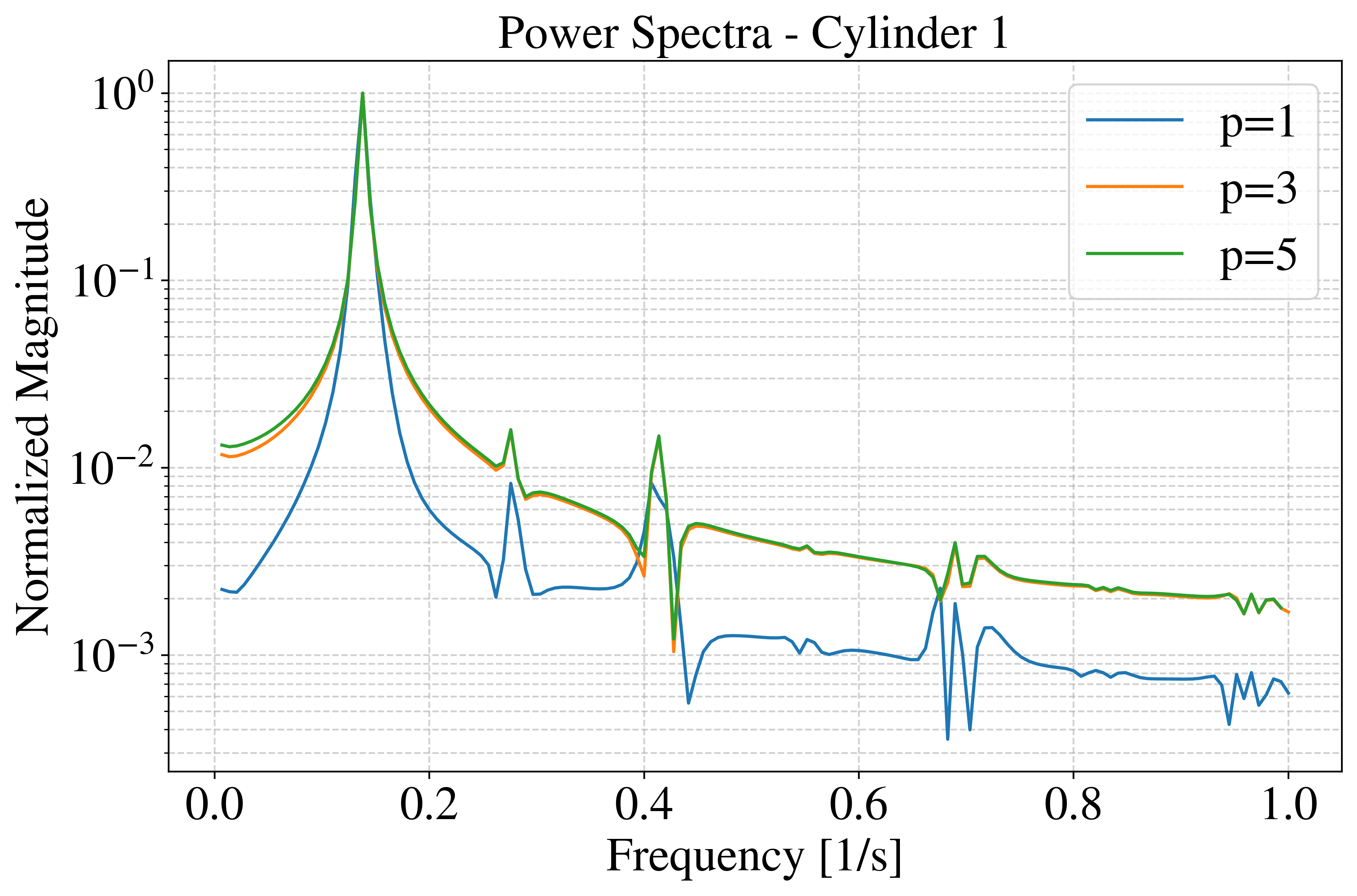}
	\hspace{\figsep}
	\includegraphics[width=0.40\textwidth]{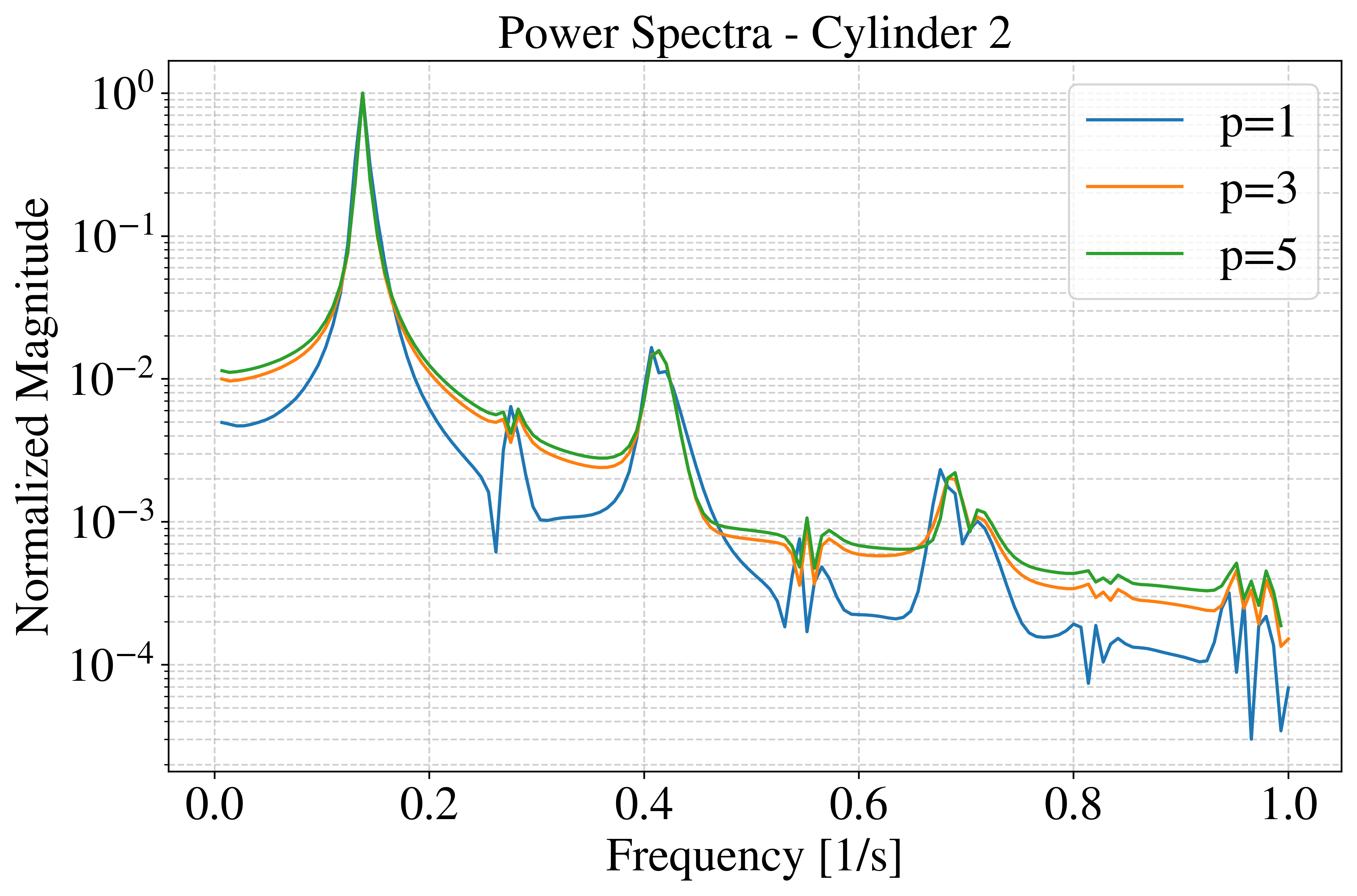}

	\caption{Normalized Power Spectra on the time series of the reduced displacements $y^*_1$ and $y^*_2$, for the two cylinder case with $U^*=7$.}
	\label{figs:insens_plot}
\end{figure}

\begin{figure}[h]
	\centering
	\includegraphics[width=0.85\textwidth,keepaspectratio]{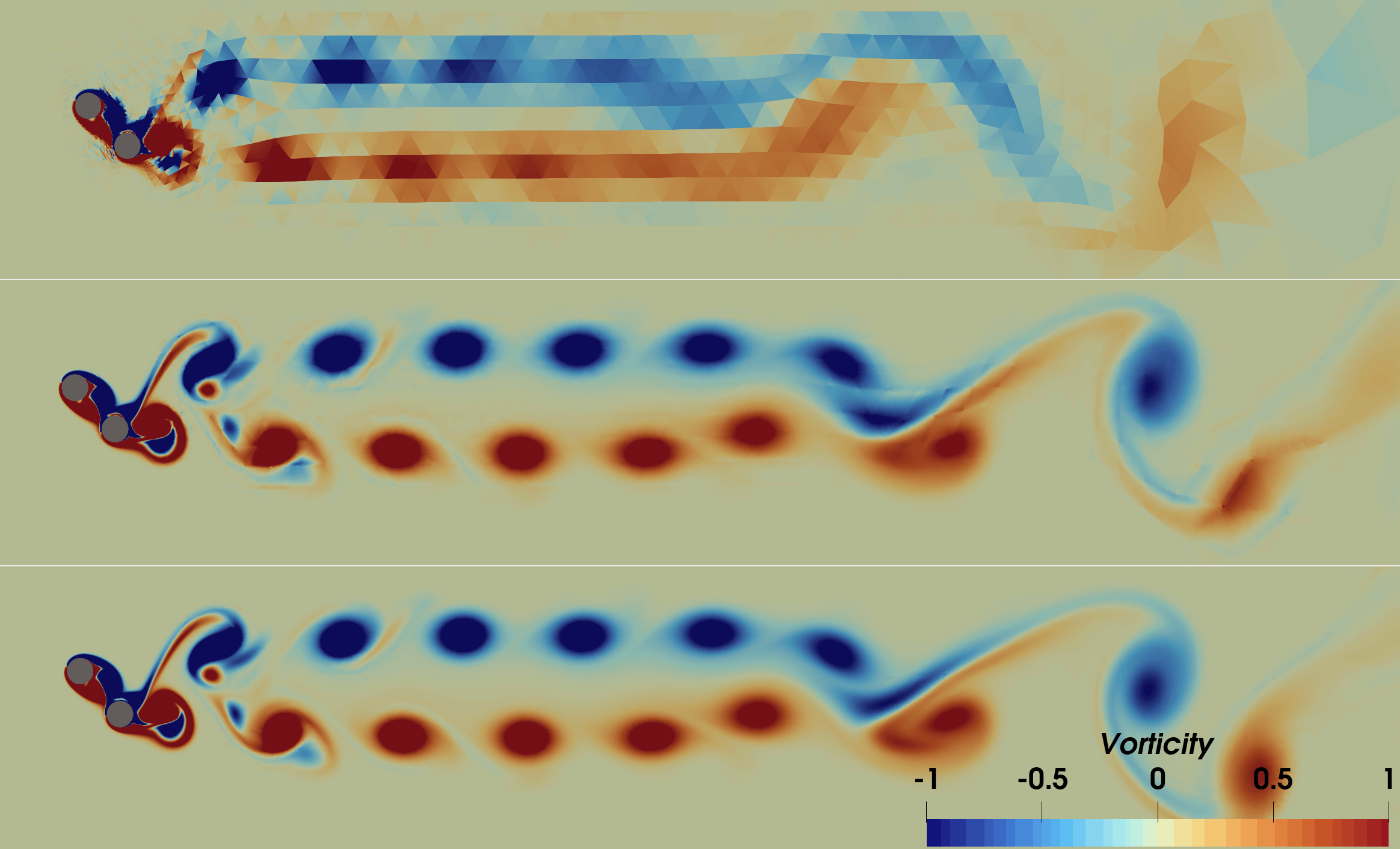}
	\caption{Vorticity Contours for $U^* = 7$ at $\sim 146 s$ for polynomial order of $p=1,3,5$. The top image is $p=1$ the middle image is $p=3$ and the bottom image is $p=5$.} \label{figs:2_cyl_vort_insens}
\end{figure}

The impact of $p$-refinement on the flow physics can be seen by the vorticity contours in Figure \ref{figs:2_cyl_vort_insens}. For $p=1$, the wake resolution is significantly degraded due to numerical diffusion, leading to a loss of fine-scale vortical structures. Conversely, the $p=3$ and $p=5$ cases exhibit exceptional wake retention.
Despite a relatively coarse mesh size in the wake region ($h \sim D$), the high-order DG method maintains the integrity of the positive and negative vortex pairs over long distances, which is a clear benefit of the method. Given the minimal accuracy gains observed between $p=3$ and $p=5$ relative to the increased computational cost, $p=3$ is selected as the optimal order for the remainder of this study.

\boldparagraph{Lissajous curves}

\begin{figure}[h]
	\centering
	\includegraphics[width=0.9\textwidth,keepaspectratio]{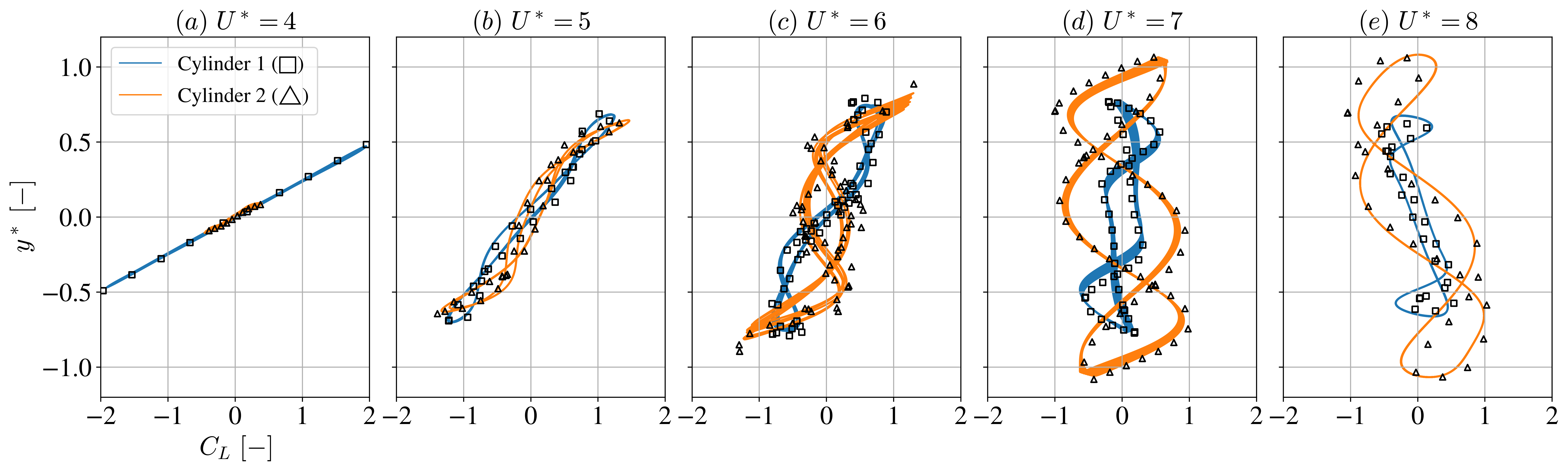}
	\caption{Plots of the Lissajous curves for the two cylinder case. The $\square$ markers refer to the results from Griffith et al. \cite{griffith2017flow} for the front cylinder and the $\triangle$ refers to the rear cylinder.}\label{figs:2_cyl_CL_y}
\end{figure}

The dynamic response of the tandem cylinders across the reduced velocity range $U^* = 4 - 8$ is captured through Lissajous phase portraits. The Lissajous phase portraits illustrate the synchronization between the instantaneous lift force and the cylinder displacement. The converged Lissajous curves for the reduced displacement $y^*$ and the lift coefficient $C_L$ are given in Figure \ref{figs:2_cyl_CL_y}.

At $U^*=4$, the strictly linear trajectory indicates that the fluid loading and structural response are perfectly in phase. As the reduced velocity increases to $U^*=5$ and $6$, the system enters a transitional regime characterized by highly irregular, quasi-periodic phase orbits. At higher reduced velocities ($U^*=7, 8$), strong periodic behavior is reestablished. The ALE-DG framework captures these highly non-linear phase dynamics in strict accordance with the literature. To gain insight about the physical mechanisms driving these phase transitions, the underlying wake topologies are investigated in the following.

\begin{figure}[h]
	\centering
	\includegraphics[width=0.85\textwidth,keepaspectratio]{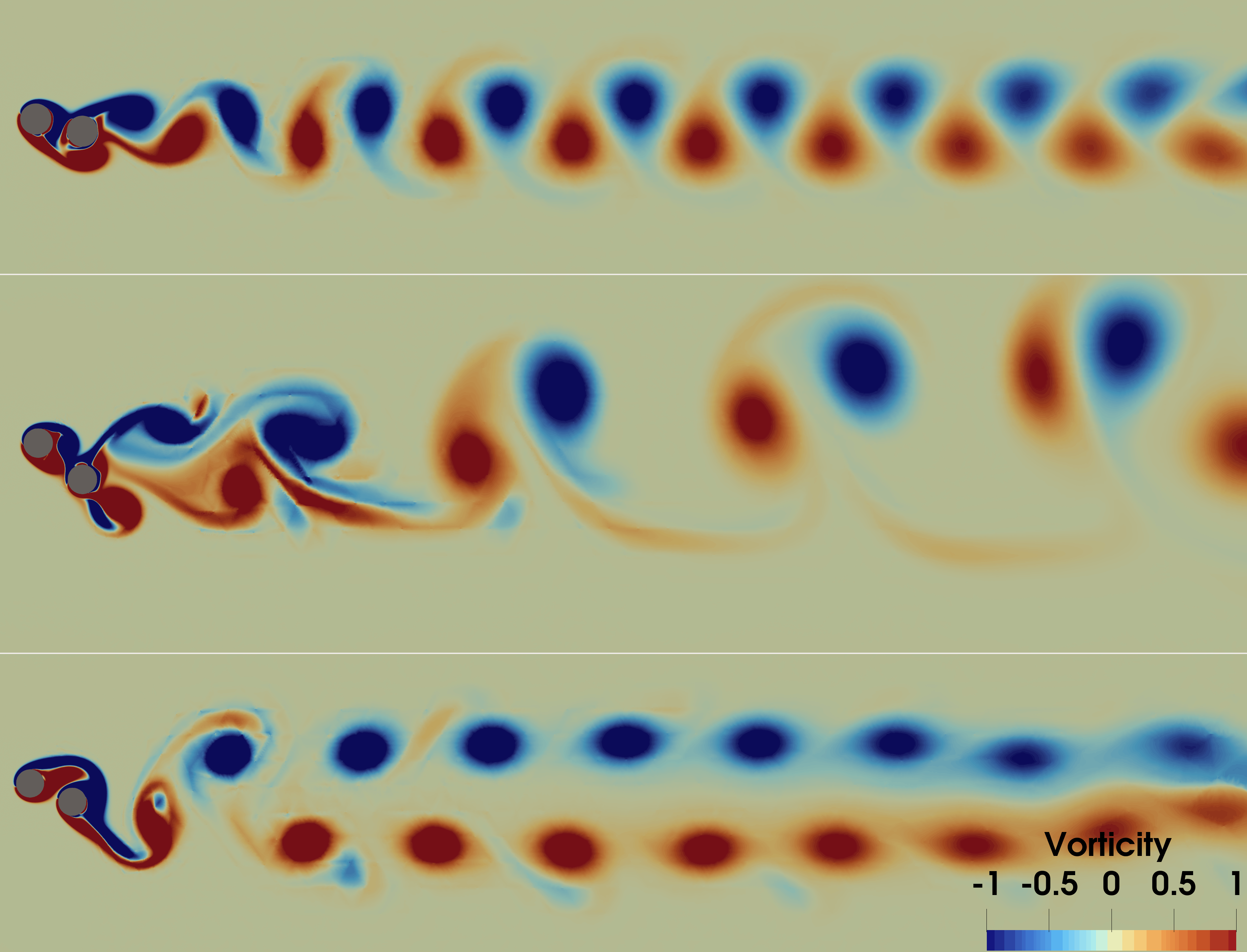}
	\caption{The three Modes of vortex shedding observed in the specific range of reduced velocities. Mode 1 is demonstrated by $U^*=4$ (top), Mode 2 by $U^*=5$ (middle) and Mode 3 by $U^*=8$ (bottom).}\label{figs:u4vsu5vsu8_2_cyls}
\end{figure}

\boldparagraph{Vortex Shedding}
The topology of the downstream wake can be found in Figure \ref{figs:u4vsu5vsu8_2_cyls}. The vortex shedding structures as per Griffith et al. \cite{griffith2017flow}, follows three modes.
Mode 1 is the $2S_R$ vortex shedding pattern with the $R$ noting that the vortices are shed from the rear cylinder. The in-between mode, Mode 2, is the $2P_F$ characterized by a pair of vortices with the $F$ noting that the vortices are shed from the front cylinder.
Lastly, Mode 3, the $2P$ mode is characterized by a pair of vortices shed by both cylinders with no distinction between the front or rear.
With this mode, wake is governed by a complex interaction where the vortex formation on the rear cylinder absorbs both the positive and negative eddies shed from the upstream body during each half-cycle \cite{griffith2017flow}. This mechanism is clearly captured in the present simulations (Figure \ref{figs:u4vsu5vsu8_2_cyls}), where the positive (red) vortical structures actively engulf the negative (blue) ones. Overall, the resolved wake topologies demonstrate very good agreement with the findings of \cite{griffith2017flow,papadakis2022hybrid}.

\begin{figure}[h]
	\centering
	\includegraphics[width=1.0\textwidth,keepaspectratio]{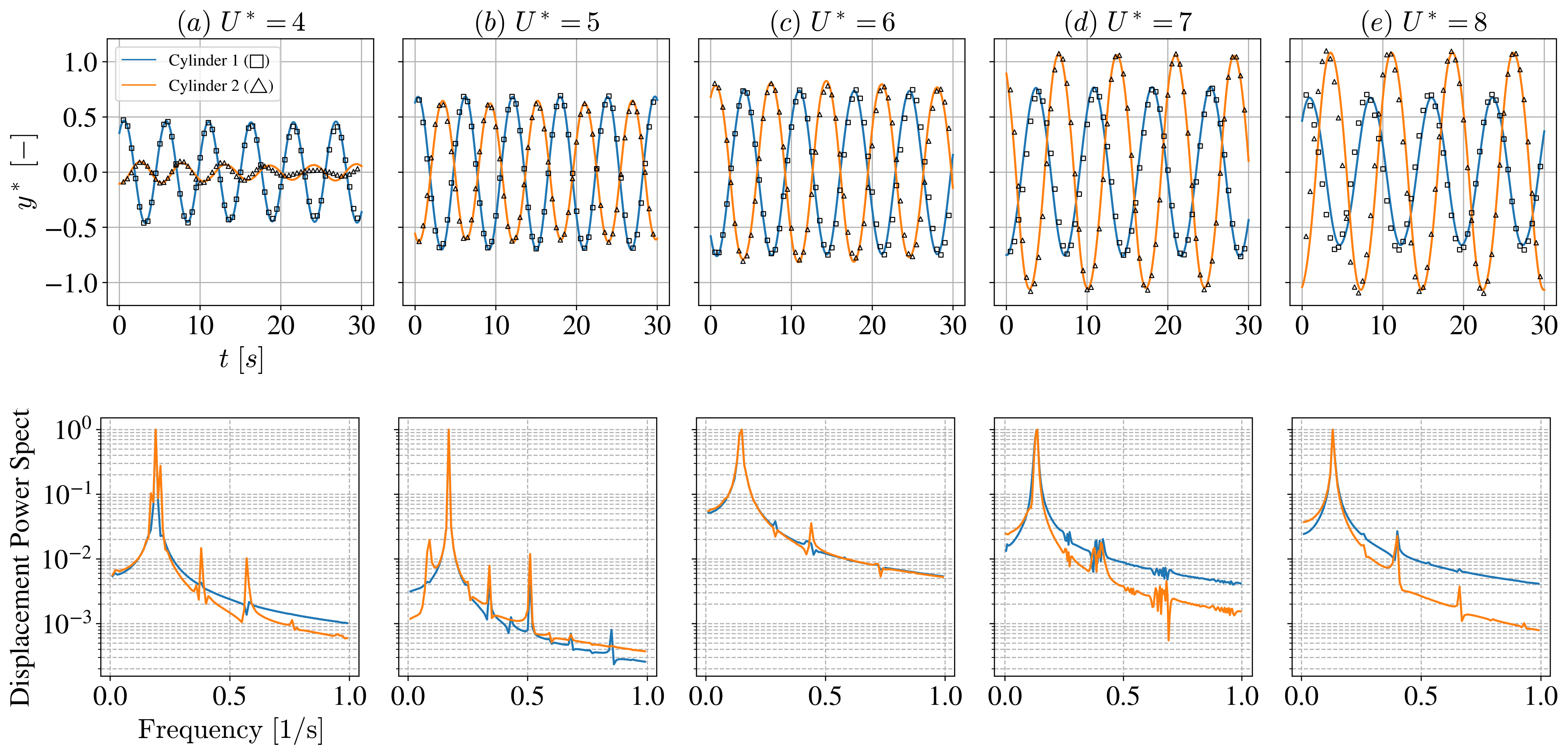}
	\caption{Time Series and Power Spectra results for the two cylinder case. The orange line ($\triangle$) refer to the rear cylinder while the blue line ($\square$) refer to the front cylinder. The markers on the time series compare the results with Papadakis et al. \cite{papadakis2022hybrid}.}\label{figs:2_cyl_ytime_fft}
\end{figure}

\boldparagraph{Time series and power spectra analysis}
Figure \ref{figs:2_cyl_ytime_fft} presents the time series of the cylinders' displacement alongside corresponding power density spectra. A direct comparison is made with data from the study of Papadakis et al \cite{papadakis2022hybrid}. For the comparison an arbitrary temporal window of 30 seconds is selected, and a constant phase shift is applied to align the signals. Across the simulated range, the responses exhibit nearly harmonic oscillations, though slight amplitude modulations are evident at $U^*=4$ and $U^*=6$. Specifically, the spectral analysis for these two cases reveals secondary frequency peaks in close proximity to the primary harmonic, which drives the observed modulations. This secondary frequency is more pronounced for the rear cylinder at $U^*=4$, also evident in the time series of \cite{papadakis2022hybrid}. Apart from this, the Fourier analysis reveals a main dominant frequency for each reduced velocity. As it can be seen in Figure \ref{figs:2_cyl_fn_fmax}, the dominant frequencies for both the upstream and downstream cylinders remain virtually identical, indicating a robust wake-induced synchronization between the two bodies. Finally, it is noted that overall the results demonstrate excellent quantitative agreement between the proposed ALE RK DG method and the literature.

\setlength{\figsep}{0.05\textwidth} 

\begin{figure}[h]
	\centering
	\includegraphics[width=0.30\textwidth]{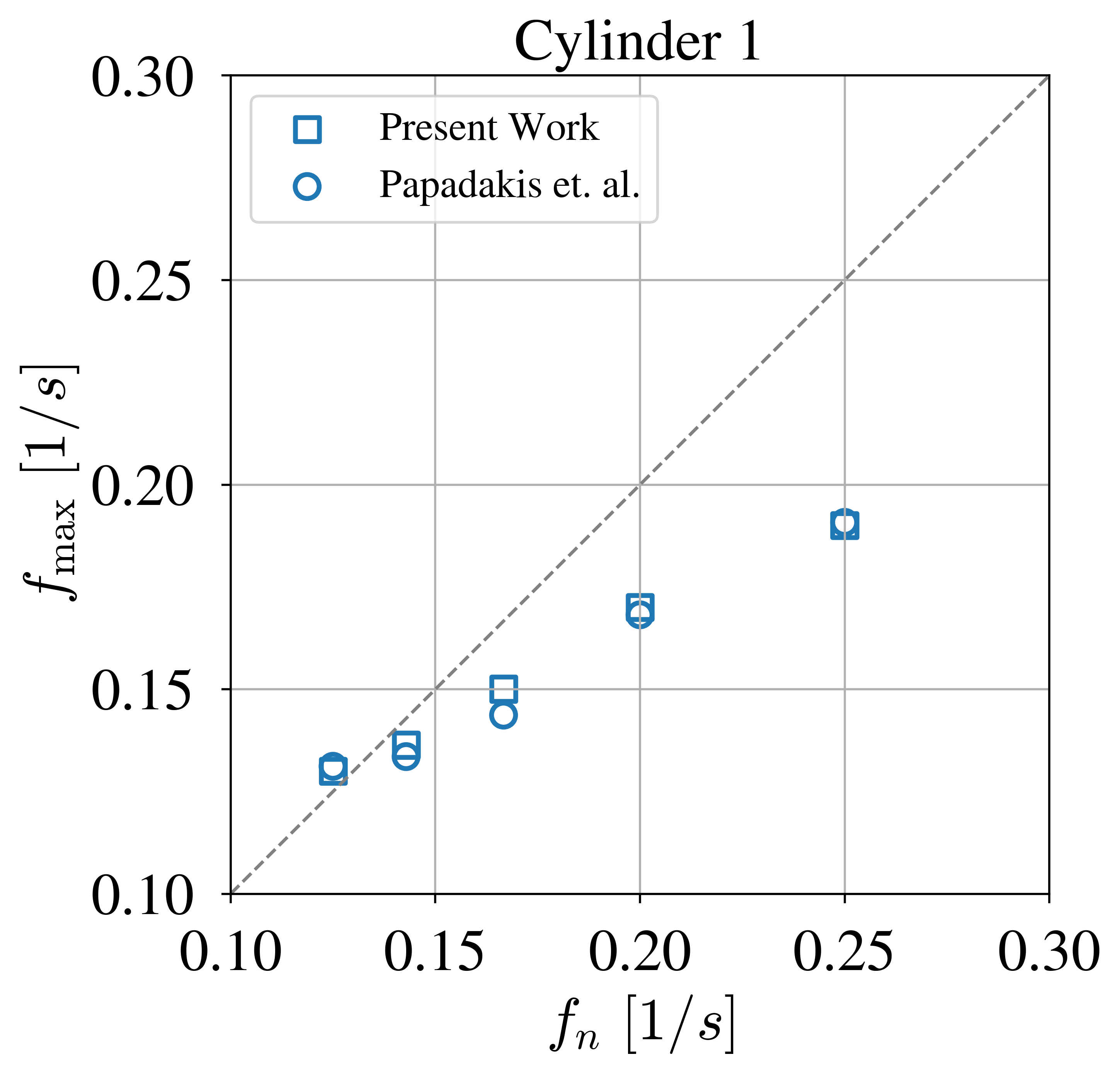}
	\hspace{\figsep}
	\includegraphics[width=0.30\textwidth]{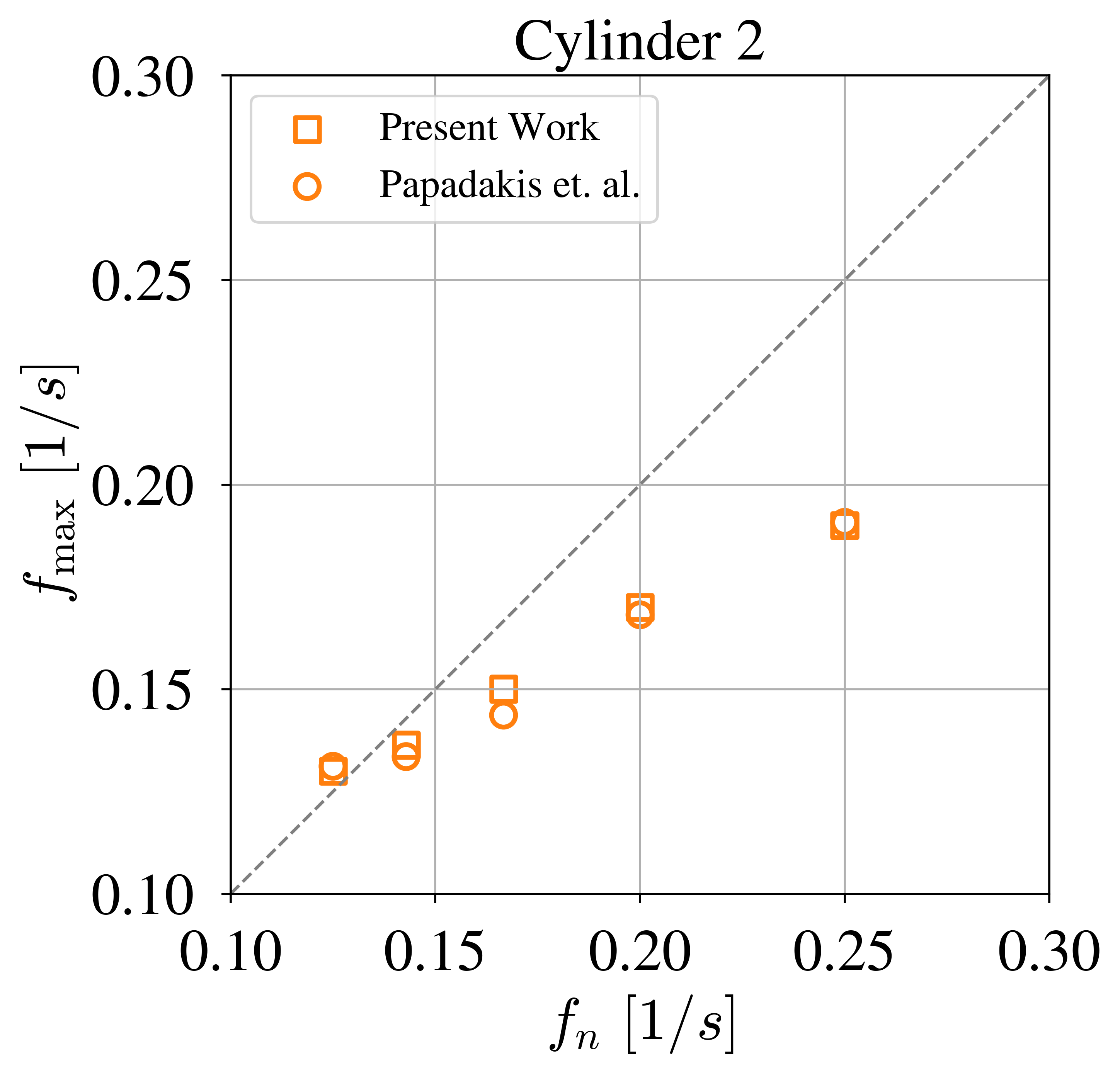}

	\caption{Frequencies corresponding to the peak of the power spectra for the two-cylinder case, with the $\square$ markers referring to the results of Papadakis et al.\ \cite{papadakis2022hybrid}. The left image (blue) refers to the front cylinder and the right image (orange) refers to the rear cylinder.}
	\label{figs:2_cyl_fn_fmax}
\end{figure}

In summary, the numerical framework replicates the dynamics of the two-cylinder tandem arrangement consistently with other published results despite the relatively coarse mesh. The vortical structures are very well preserved in the wake with negligible numerical dissipation, and the time series, power spectra, and Lissajous curves closely match those reported in the literature, demonstrating the solver's ability to capture both amplitude and phase of the cylinder oscillations. Furthermore, even under the large structural deformations encountered in this configuration, the solution remains largely unaffected, indicating strong robustness of the formulation.

\subsection{Three Cylinders in Tandem Arrangement}
Following the 1-DoF two-cylinder system, we examine a more demanding fluid-structure interaction scenario: three tandem cylinders possessing both stream-wise and cross-flow kinematic freedom (2-DoF per body).
This system exhibits highly non-linear coupled dynamics, as the independent trajectories of the bodies are heavily dictated by the wake interactions.
Consequently, small differences in the resolved vortical structures can result in significant deviations in the predicted response.
To evaluate the ALE DG solver's capability in this more complex scenario our computational results are compared against the literature, specifically the work of Yu et al \cite{Yu2016FlowInduced}.
It is noted here that in \cite{Yu2016FlowInduced} they employed a high-order FEM approach.

\begin{figure}[h]
	\centering
	\includegraphics[height=0.15\textwidth,keepaspectratio]{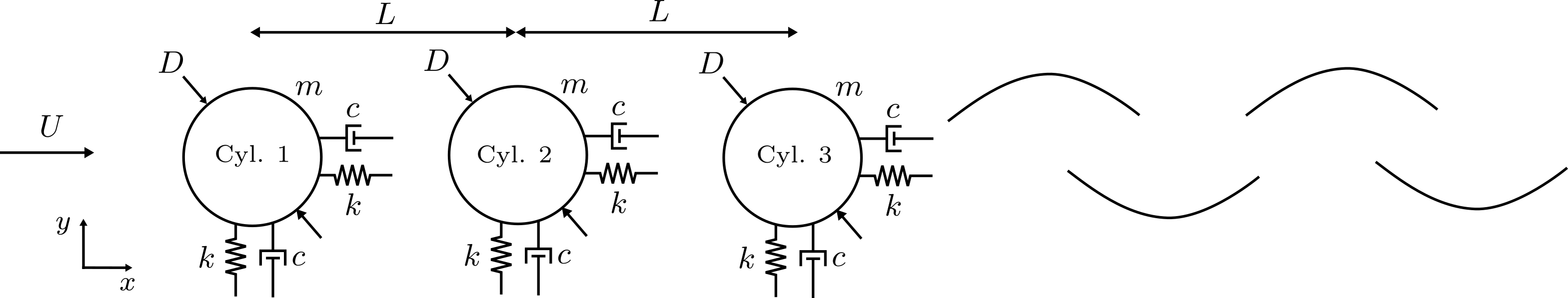}
	\caption{The three cylinders in tandem arrangement with 2 DoF in both in-flow and cross-flow directions.
	}\label{figs:3_cyl_config}
\end{figure}

The computational setup, illustrated in Figure \ref{figs:3_cyl_config}, consists of three equally spaced cylinders with a center-to-center distance of $L/D = 4$. The dynamic system's properties, including spring stiffness and damping, are the same across both spatial axes. Also $m^* = m/m_f = 4/\pi$. The reduced displacements are also defined as $x^* = x / D$, $y^* = y / D$. Simulations are conducted at a fixed Reynolds number of $\mathrm{Re} = 150$, ensuring the flow remains strictly within the laminar regime and no 3D fluid effects should, in principle, occur.

\begin{figure}[h]
	\centering
	\includegraphics[height=0.35\textwidth,keepaspectratio]{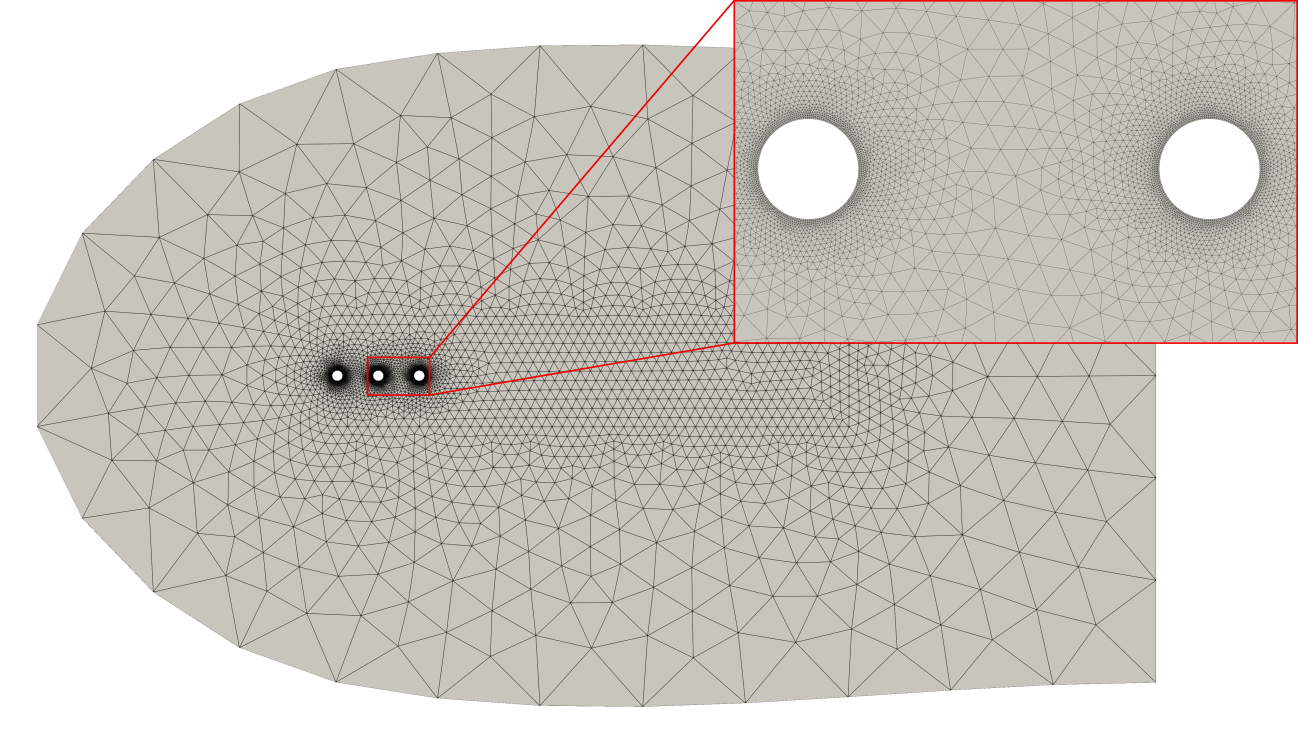}
	\caption{Computational grid for the three cylinder in tandem case.}\label{figs:3_cyl_grid_sparse}
\end{figure}

The computational domain is discretized using approximately $14,700$ elements, while preserving a surface resolution of 160 nodes per cylinder, consistent with the previous configuration. The resulting mesh is presented in Figure~\ref{figs:3_cyl_grid_sparse}, together with a magnified view of the grid in the vicinity of the cylinders. The local grid topology remains identical to that of the previous case, with the only modification being the increased spacing between the cylinders.

\begin{figure}[h]
	\centering
	\includegraphics[width=0.45\textwidth,keepaspectratio]{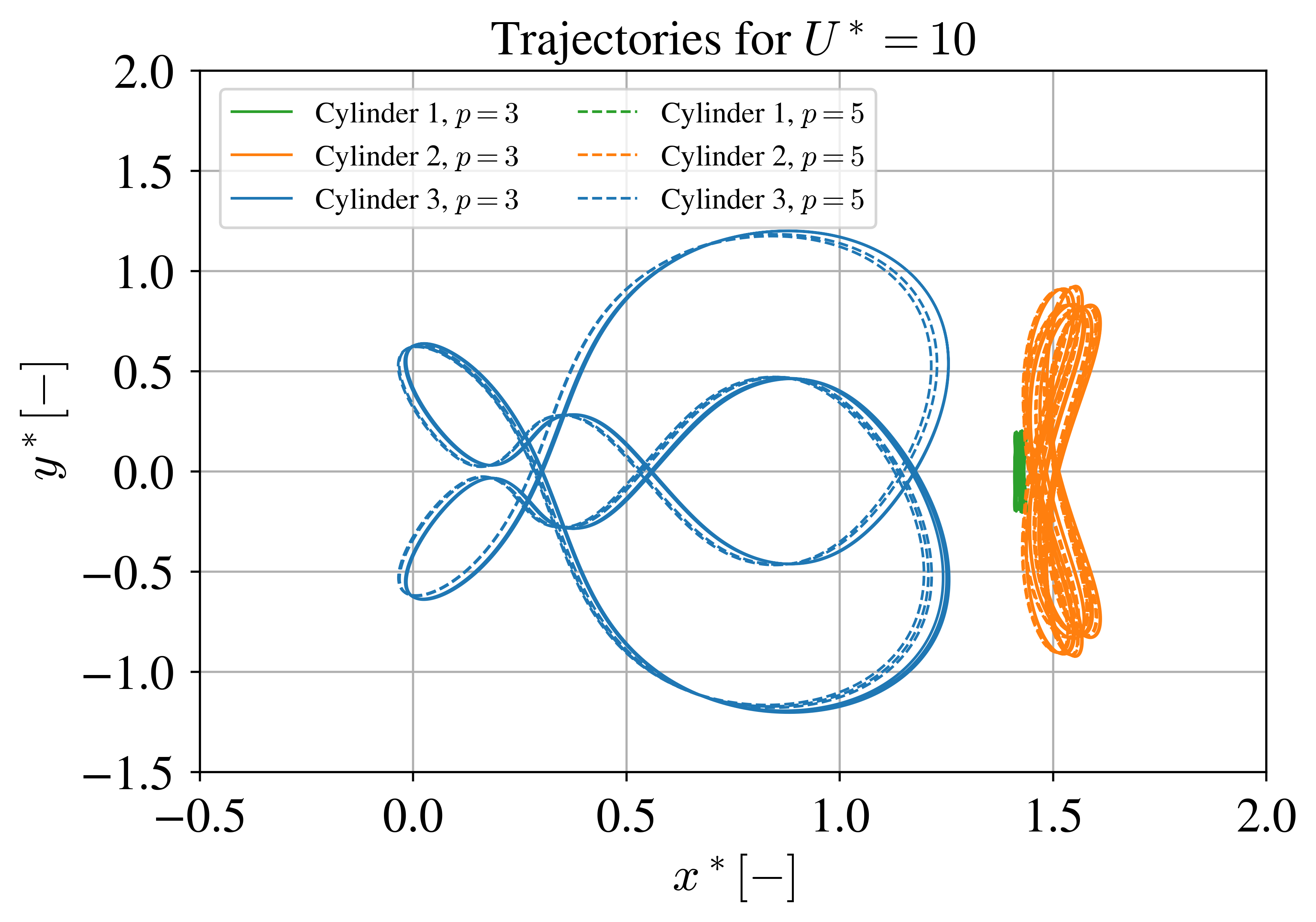}
	\caption{Cylinder trajectories results for the three cylinder case, for $U^*=10$. The green, orange and blue curves refer to a polynomial order of 3 while the red, purple and brown to a polynomial order of 5.}\label{3_cyl_trejactories_p3p5}
\end{figure}

The dynamic response of the system is evaluated across a range of reduced velocities, $U^* \in [5, 10]$.
To verify the  convergence of the ALE-DG framework, a $p$-refinement study ($p=3$ versus $p=5$) is conducted specifically for $U^*=10$. This specific reduced velocity was chosen due to the complex trajectories of the back and middle cylinder as well as the large amplitudes in the cross-flow and in-flow directions of the trailing cylinder. Furthermore, $U^*=10$ is characterized by strongly periodic multi-body interactions, providing a more reliable baseline for comparison, as opposed to the  irregular trajectories observed at lower velocities. As illustrated in Figure \ref{3_cyl_trejactories_p3p5}, both polynomial orders yield remarkably similar  trajectories, with only a marginal phase shift distinguishing the two. Given the substantial increase in computational overhead associated with the $p=5$ discretization, $p=3$ is used for all subsequent simulations.

\begin{figure}[h]
	\centering
	\includegraphics[width=0.98\textwidth,keepaspectratio]{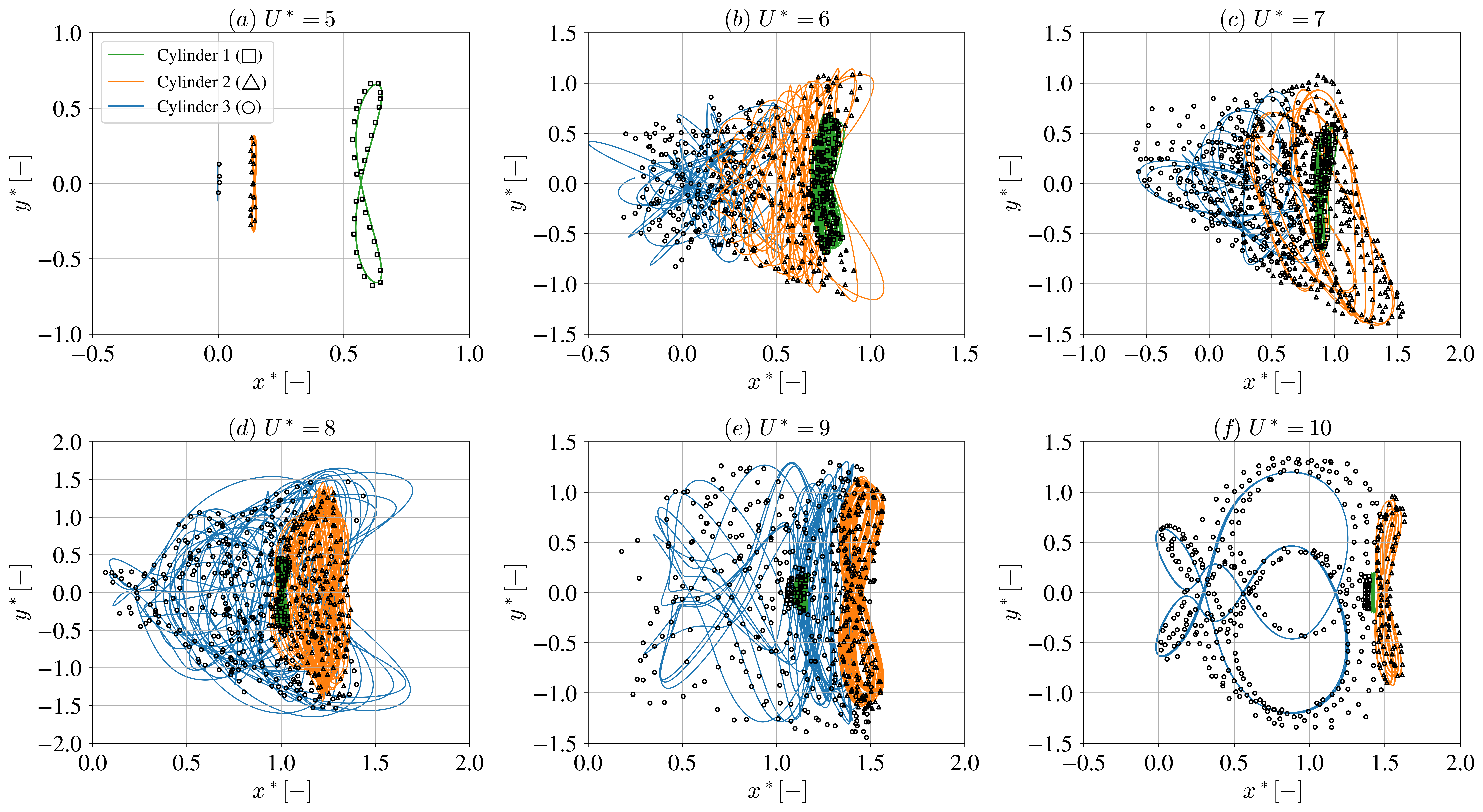}
	\caption{Cylinder Trajectories for the three cylinder case with different reduced velocities $U^*$. The markers refer to the results from Yu et al. \cite{Yu2016FlowInduced} with $\triangle$ marker referring to the back cylinder, $\square$ marker referring to the middle cylinder and $\bigcirc$ marker referring to the rear cylinder. }\label{3_cyl_trejactories}
\end{figure}

\boldparagraph{Cylinder Trajectories}
The response of the three-cylinder system is examined across the reduced velocity range $U^* \in [5, 10]$, with the resulting  trajectories presented in Figure \ref{3_cyl_trejactories} alongside the data of Yu et al. \cite{Yu2016FlowInduced}. At $U^*=5$ the system responds with an ordered periodic motion. The downstream cylinder traces a distinct figure-eight pattern, while the upstream and intermediate cylinders show similar shapes but are more confined. As the reduced velocity increases to $U^*=6$ and $7$, the system transitions into a highly irregular regime driven by complex wake interactions. In these two cases, the downstream cylinder's response does not follow a specific pattern, while the upstream body develops a clear asymmetric figure-eight orbit. At higher velocities, $U^*=8-10$, the downstream cylinder becomes highly ``energized'', undergoing large-amplitude motion in both the stream-wise and cross-flow directions. Lastly, at $U^*=10$ distinct periodicity re-emerges and a characteristic ``butterfly'' trajectory is observed.

Throughout the entire reduced velocity range, the ALE-DG solver demonstrates good qualitative agreement with the \cite{Yu2016FlowInduced} even though, the complexity of the cylinders trajectories increases  at higher reduced velocities. Most deviations with the results of \cite{Yu2016FlowInduced} occur in the downstream cylinder while the agreement for the other two is very good. Such  discrepancies are expected, given the highly irregular and sensitive wake interactions governing the rear body's motion.

\begin{figure}[h]
	\centering
	\includegraphics[width=0.98\textwidth,keepaspectratio]{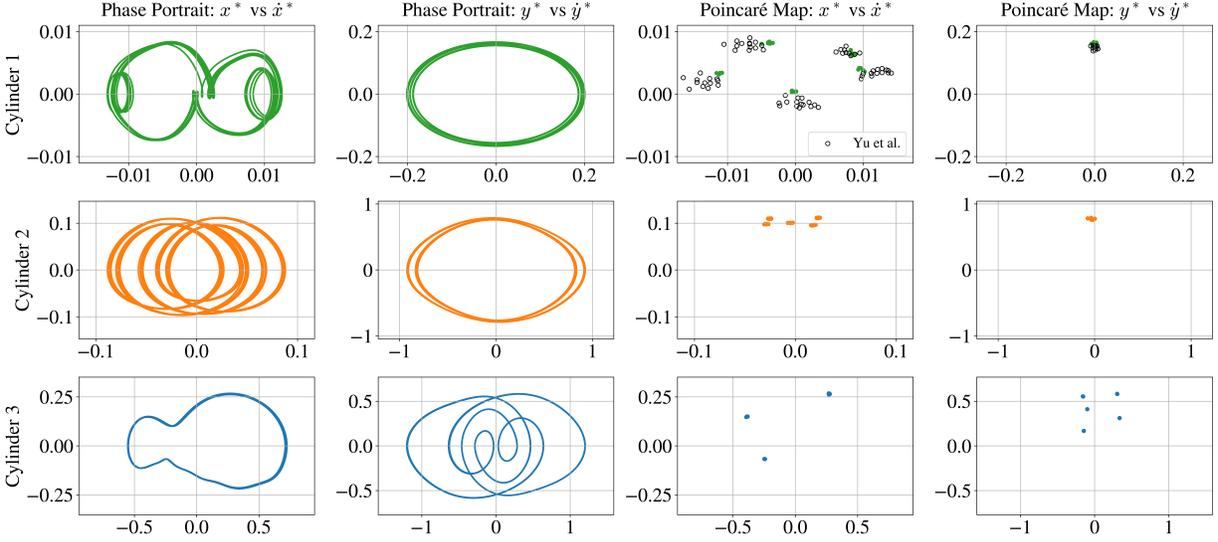}
	\caption{Phase Portraits and Poincaré Maps for the three cylinder case, with a reduced velocity of $U^*=10$. Whenever they are available the reference points from Yu et al. \cite{Yu2016FlowInduced} are included plotted using the $\bigcirc$ marker.}\label{3_cyl_poincare_U10}
\end{figure}

\boldparagraph{Phase Portraits and Poincaré Maps}
To further characterize the system's non-linear behavior, phase portraits and their corresponding Poincaré maps are constructed following the methodology of Yu et al. \cite{Yu2016FlowInduced}. The phase-space dynamics for the $U^* = 10$ regime can be found in Figure \ref{3_cyl_poincare_U10}. For the upstream and middle bodies, cylinders 1 and 2, the cross-flow ($y$) response is periodic, characterized by a single point in the Poincaré map. Conversely, the stream-wise ($x$) responses (and $y$ for the downstream cylinder)  show quasi-periodicity with the points clamping around several areas in the map.
Furthermore, the overlapping loops present in the middle cylinder's stream-wise phase portrait correspond directly to the figure-eight spatial trajectories observed previously. The complex shapes indicate that the dynamics of the system are heavily irregular, but with some quasi-periodicity being present.
The Poincaré maps align well with the bibliography \cite{Yu2016FlowInduced} for the up-stream cylinder with the stream-wise response being clamped around five points and the cross-stream response being clamped around a single point.

\begin{figure}[h]
	\centering
	\includegraphics[width=0.98\textwidth,keepaspectratio]{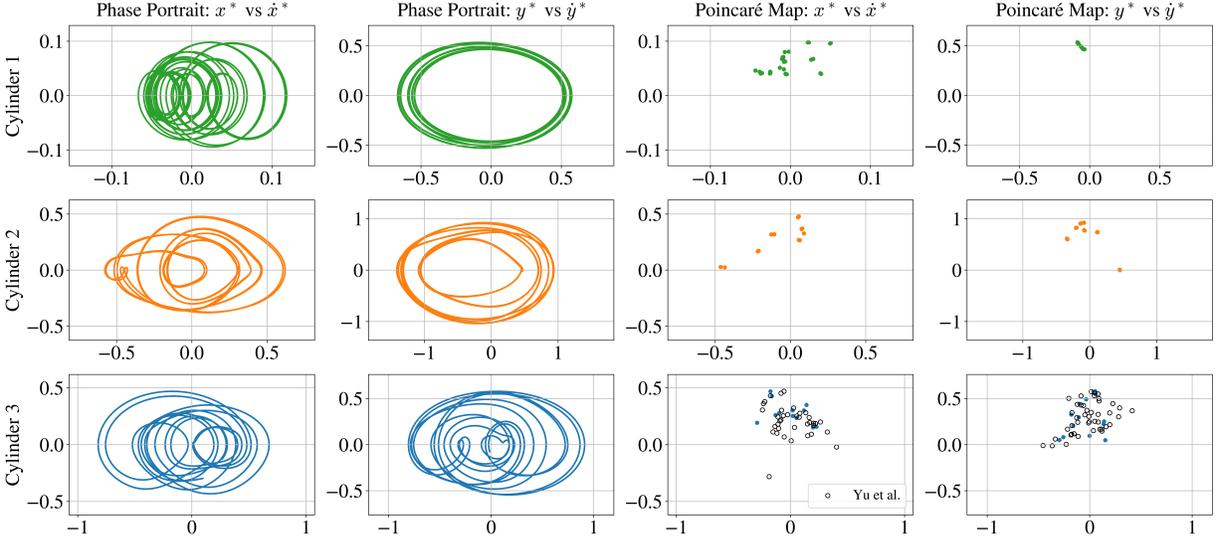}
	\caption{Poincaré Phase Portraits and Maps for the three cylinder case, with a reduced velocity of $U^*=7$. Whenever they are available the reference points from Yu et al. \cite{Yu2016FlowInduced} are included plotted using the $\bigcirc$ marker.}\label{3_cyl_poincare_U7}
\end{figure}

A highly irregular case is the one of $U^*=7$, as seen in Figure \ref{3_cyl_poincare_U7}, with the points scattering over the plane for most cylinders instead of gathering around specific points. Similar patterns are observed for the cases of $U^*=6,8,9$ but they are omitted. This makes the dynamics for these reduced velocities very complex and thus no pattern can be observed for such cases. A stable pattern that does however occur, is the periodic nature of the $y$-oscillation of cylinder 1, with its orbit being circular. This leads to the conclusion that, although the dynamics of the back and middle cylinder become sporadic the front cylinder remains more stable.
For the downstream cylinder, the contour maps show good agreement with the literature \cite{Yu2016FlowInduced}, with the data points distributed across the plane in a similar manner.

\begin{figure}[h]
	\centering
	\includegraphics[width=0.85\textwidth,keepaspectratio]{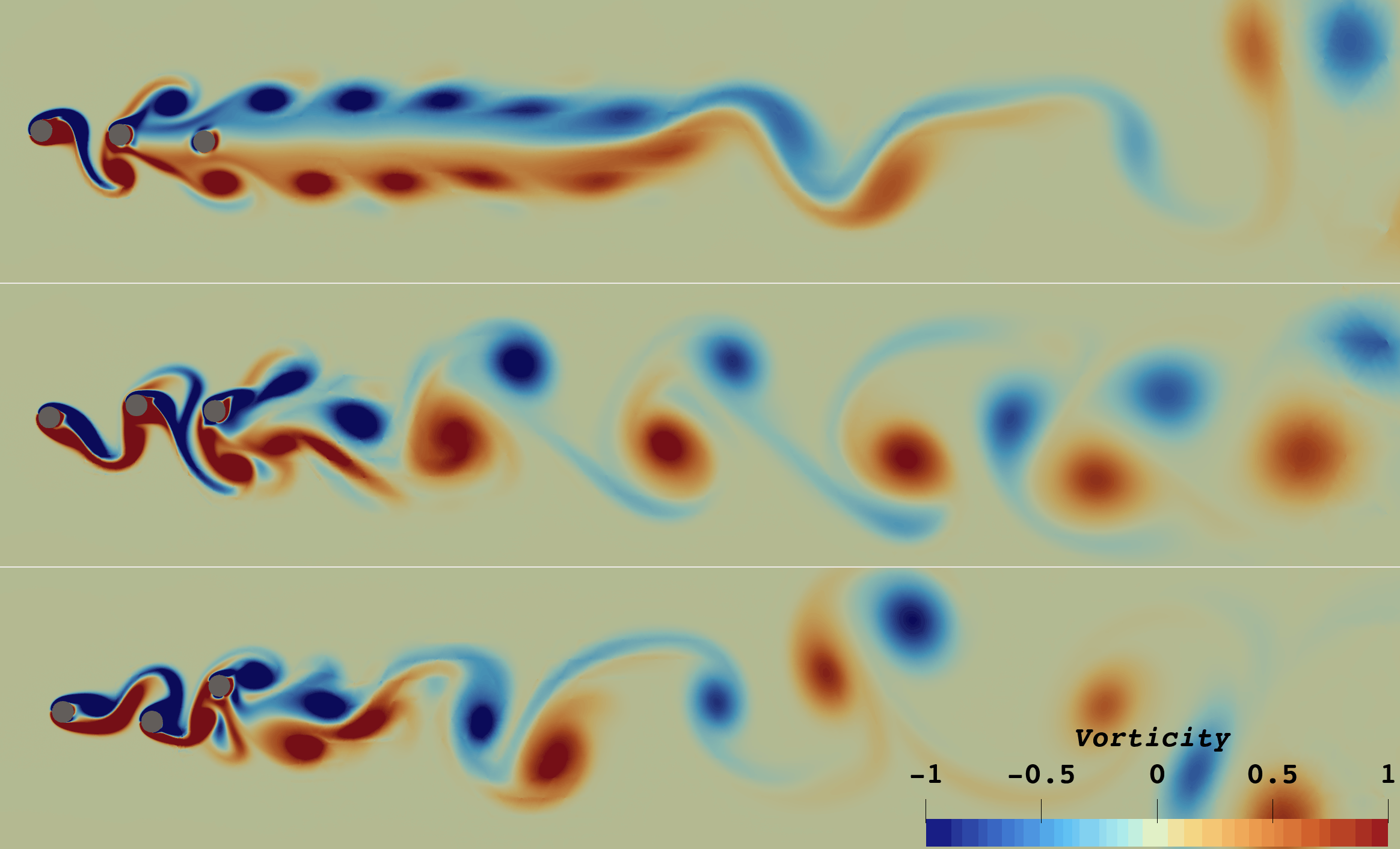}
	\caption{Vortex shedding for the $U^* = 5,7,10$ cases, with the top representing a reduced velocity of 5, the middle a reduced velocity of 7 and the bottom a reduced velocity of 10.}\label{3_cyl_vort}
\end{figure}

\boldparagraph{Vortex Shedding}
The wake topology for three reduced velocities can be found in Figure \ref{3_cyl_vort}. For $U^* = 5$, the flow is characterized by two distinct, parallel vortex streets indicative of a stable co-shedding regime. This topology aligns closely with the numerical observations of Gao et al. \cite{gao2020numerical} for equivalent reduced velocities and inter-cylinder spacing.
As the reduced velocity increases, the wake transitions into a highly complex, non-linear interaction regime. Specifically, at $U^* = 7$ and $10$, the wake becomes dominated by the shedding of co-rotating vortex pairs.
This complex interaction closely corresponds to the classical ``2C'' vortex shedding mode as classified by Williamson and Govardhan \cite{williamson2004vortex}.

\begin{figure}[h]
	\centering
	\includegraphics[width=0.95\textwidth,keepaspectratio]{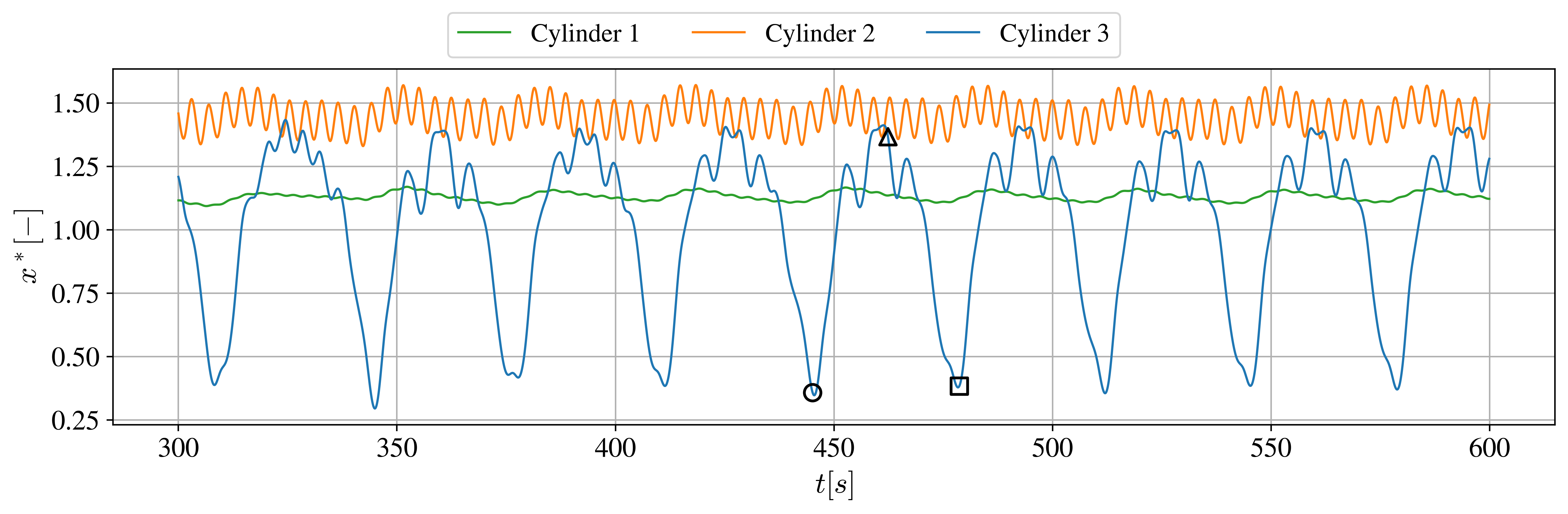}
	\caption{Time series for reduced displacement $x^*$ for $U^* = 9$. The markers refer to the time snapshots for the vortex shedding structures from Figure \ref{3_cyl_vort_attract}.}
	\label{3_cyl_time_series_attract}
\end{figure}

\begin{figure}[h]
	\centering
	\includegraphics[width=0.85\textwidth,keepaspectratio]{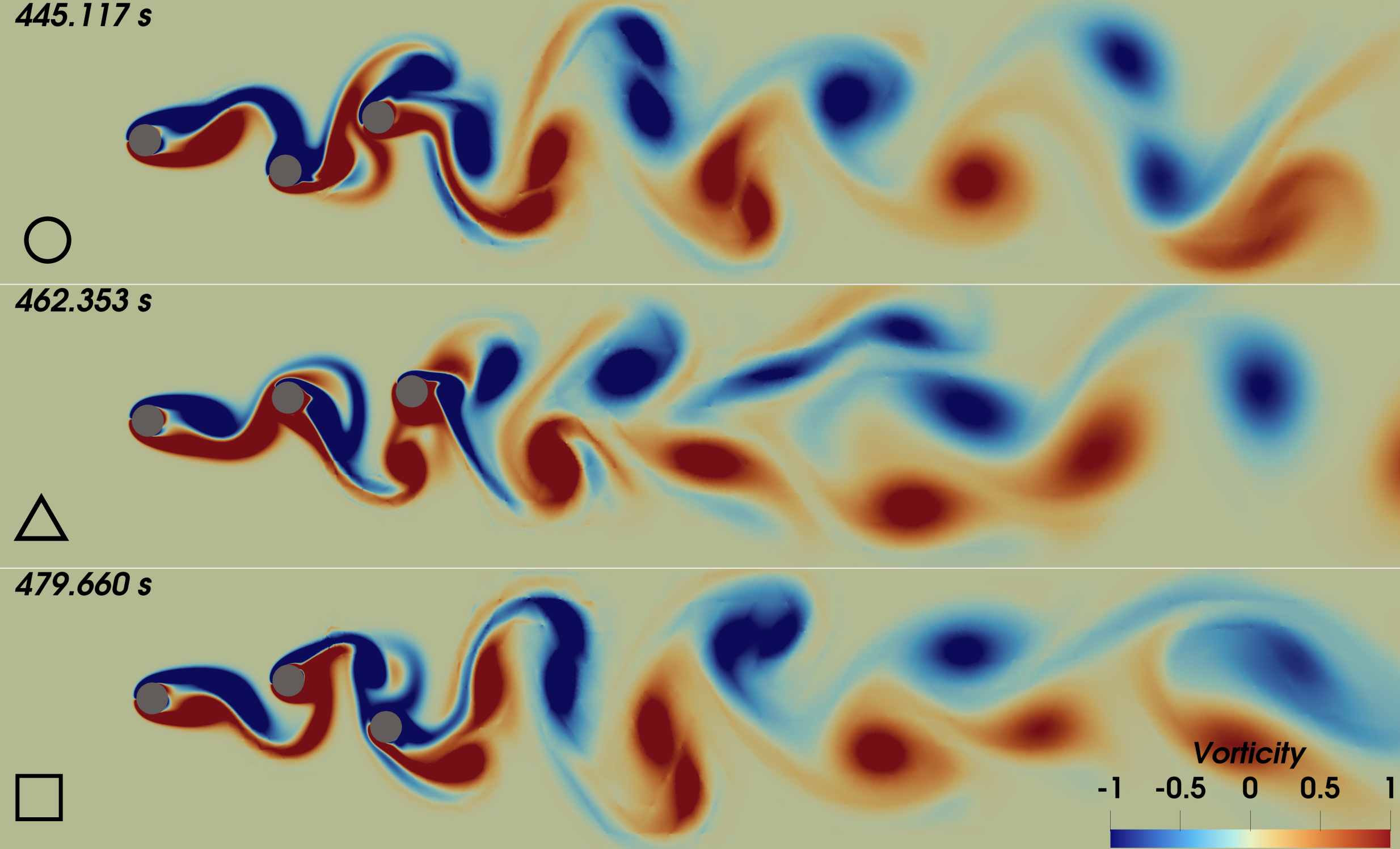}
	\caption{Vortex shedding patterns for $U^* = 9$ for different time snapshots.
	}
	\label{3_cyl_vort_attract}
\end{figure}

Another interesting feature observed in the $U^* = 9$ case is the periodic "attract-and-release" cycle characterizing the stream-wise oscillations. In Figure \ref{3_cyl_time_series_attract}, the $x$-response time series of the trailing cylinder can be found.
The $\bigcirc,\triangle$ and $\square$ symbols are there to provide visual correspondence with the vorticity snapshots of Figure \ref{3_cyl_vort_attract}.
It is evident that the cylinders exhibit periodic amplitude modulations. It is noted also that, because these amplitude drops occur periodically, they are attributed to sustained fluid-structure coupling rather than isolated disturbances.

In detail, at $t \approx 445$ s and $t \approx 479$ s, the system is in the ``release'' phase; the trailing cylinder's motion is heavily reduced, and the close spacing of the bodies prevents the shed vortices from merging. This forces the shedding of co-rotating vortices into a triplet formation (called ``pseudo-2T'' \cite{Yu2016FlowInduced}), as seen in Figure~\ref{3_cyl_vort_attract}. The emergence of this pattern severely hinders the downstream cylinder's structural response.

By contrast, at $t \approx 462$ s, the response reaches a peak (see Figure \ref{3_cyl_time_series_attract}). During this phase, the trailing cylinder vibrates more freely and is less influenced by the upstream cylinders. This increased spatial separation allows the vortices to merge. These alternating wake states and their corresponding structural impacts are in excellent agreement with the findings of Yu et al \cite{Yu2016FlowInduced}.

\begin{figure}[h]
	\centering
	\includegraphics[width=0.85\textwidth,keepaspectratio]{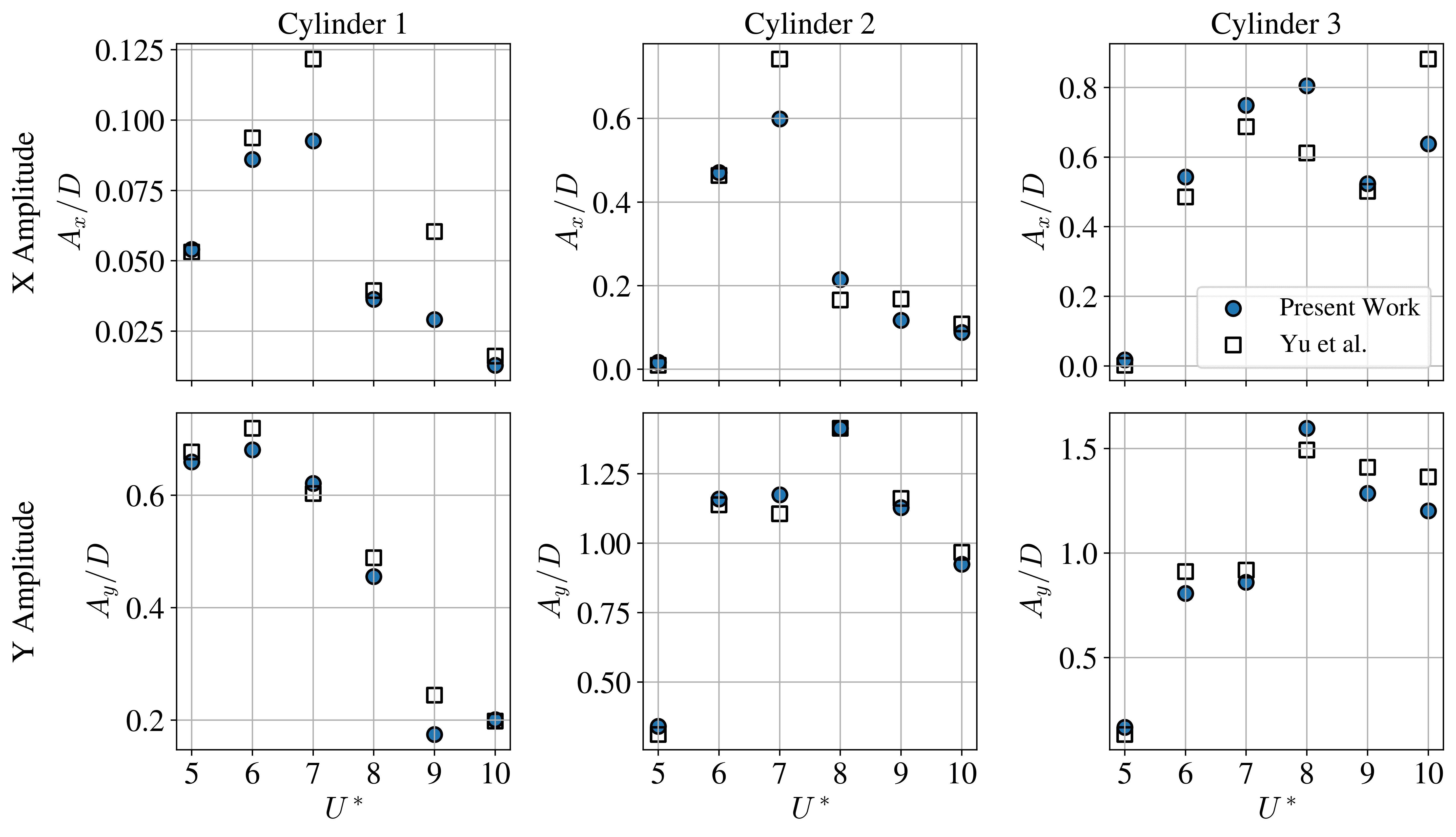}
	\caption{Maximum amplitudes computed as $A_{.} = ( \max_t(.) - \min_t(.) ) / 2$ (filled blue circles). They are compared with the maximum amplitudes from Yu et al. \cite{Yu2016FlowInduced} (hollow black squares) }\label{max_amps_3_cyl}
\end{figure}

Finally,  the amplitudes of  the oscillations are compared to those of Yu et al. \cite{Yu2016FlowInduced} as it can be seen in Figure \ref{max_amps_3_cyl}. The result from our ALE DG method for the $y$-oscillation is in agreement with the literature's data with relatively small deviations. Similarly, $x$-responses are also in good agreement, however there are some notable outliers for the reduced velocities of $U^* \ge 7$. These localized discrepancies, particularly evident in the trailing cylinder (Cylinder 3), can be attributed to the highly complex nature of the multi-body wake interactions. The previously discussed "attract-and-release" mechanism continuously disrupts the stream-wise kinematics, preventing the downstream cylinder from locking into a synchronized, stable periodic state.

For the upstream body (Cylinder 1), the overall stream-wise amplitudes are inherently small. Consequently, the large deviations observed at $U^*=7$ and $U^*=9$ actually correspond to small differences of the order of just $\Delta(A_x/D) \approx 0.02$. At $U^*=8$, the observed differences are driven by an expansion of the trajectory due to the larger excitation of the  trailing cylinder. Finally, for the $U^*=10$ case, in  \cite{Yu2016FlowInduced}  they find a wider spread in the stream-wise excursions, whereas the ALE-DG framework predicts more tightly confined, concentrated trajectories.

Overall, the proposed numerical framework is shown to be able to capture accurately the complex dynamics of the three-cylinder system. The high-order DG discretization effectively preserves the irregular and complex vortical structures, enabling a detailed and reliable resolution of the coupled cylinder motion. The ALE formulation exhibits robust performance under large structural deformations, particularly at higher reduced velocities where oscillation amplitudes approach $\sim 2D$. Moreover, the resulting vortical patterns are in close agreement with those documented in the literature, further supporting the accuracy of the method.

\subsubsection{hp-Refinement}
The preceding results were obtained using $p=3$ on a relatively coarse mesh. A natural question is whether comparable accuracy could be achieved more efficiently using a finer mesh with a lower polynomial order ($h-$refinement). To address this, we compare $h$-refinement against $p$-refinement i.e., a polynomial order increase, for the $U^* = 10$ case.

\begin{figure}[h]
	\centering
	\includegraphics[height=0.35\textwidth,keepaspectratio]{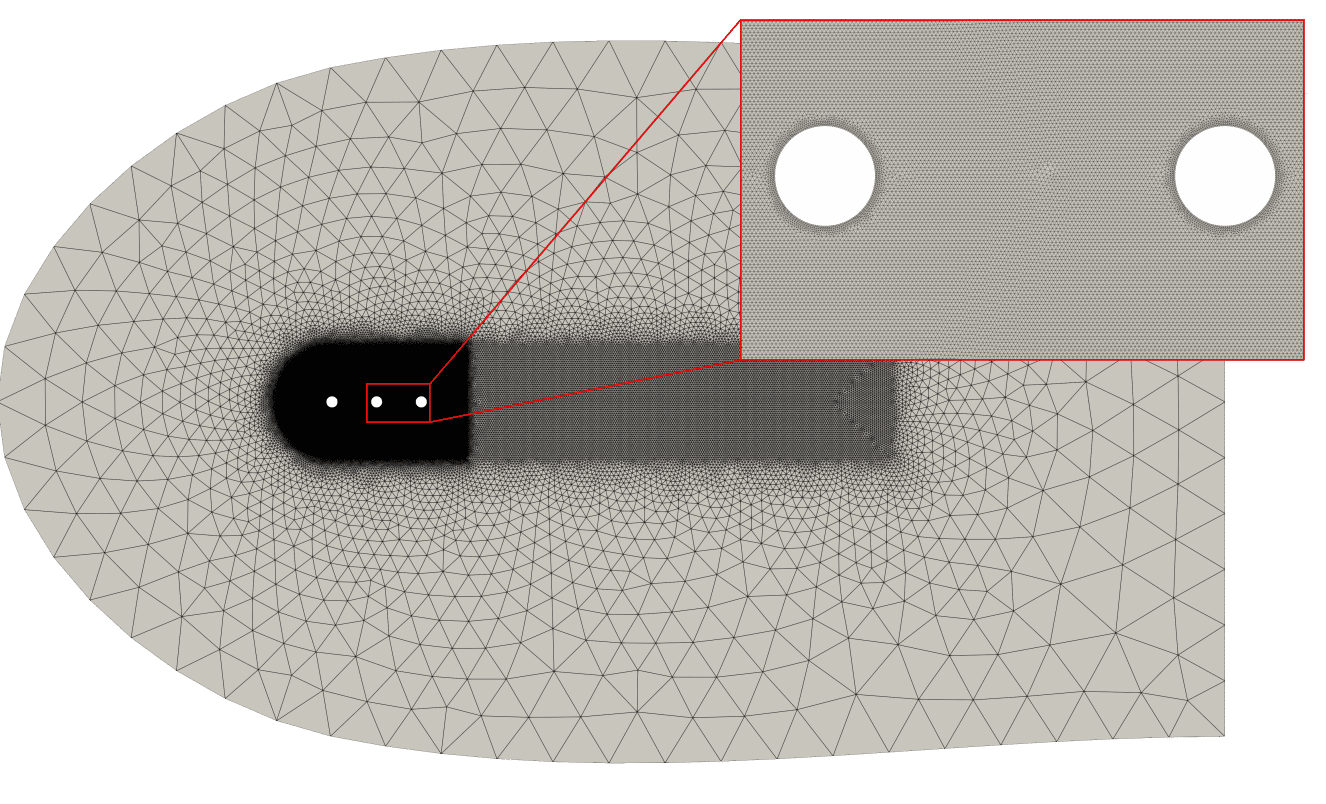}
	\caption{Dense first-order mesh for $h$-refinement for the three tandem cylinder case.}\label{figs:3_cyl_p1_dense_grid}
\end{figure}

More specifically, for the $h$-refinement, $p=1$ is used while the mesh is significantly denser, consisting of approximately $279,000$ elements and $160$ nodes on the cylinders' surface; see Figure \ref{figs:3_cyl_p1_dense_grid}, as opposed to the initial ``coarse'' mesh as shown in Figure \ref{figs:3_cyl_grid_sparse}.
More specifically, for the dense $p=1$ case, the grid close to the cylinders has an element size of $D / 25$ while at the wake region the element size is equal to $D / 5$.

\begin{figure}[h]
	\centering
	\includegraphics[width=1.0\textwidth,keepaspectratio]{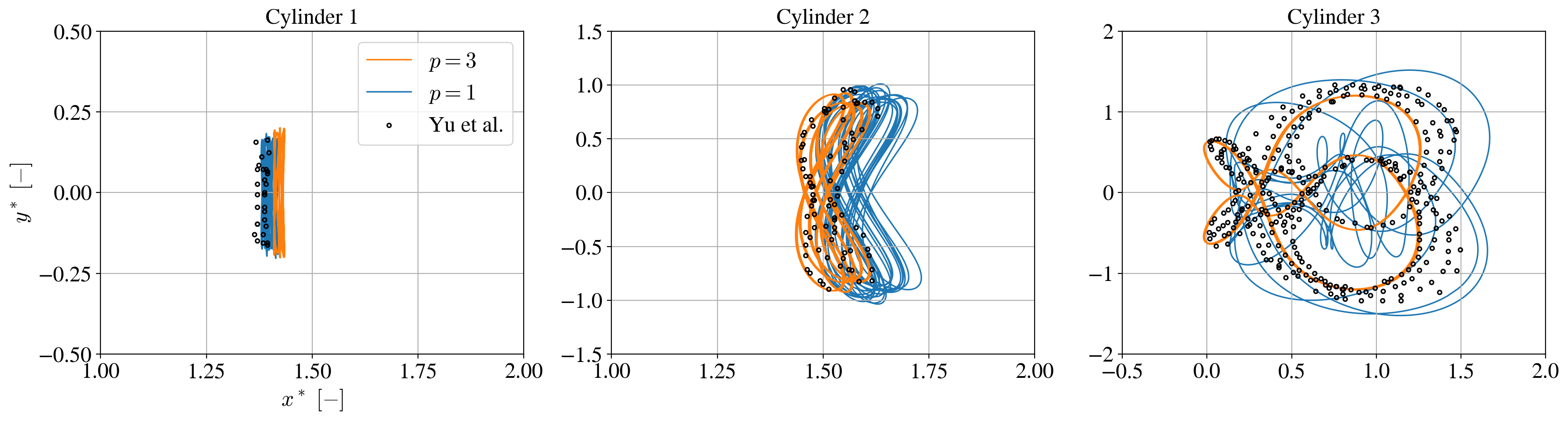}
	\caption{Comparison of cylinder trajectories for $U^* = 10$ between first-order and third-order simulations, alongside the reference data from the literature.}
	\label{fig:p1_vs_p3_3_cyl}
\end{figure}

For this comparison, we consider the $U^* = 10$ case.
The resulting trajectories for the coarse grid $p=3$ and the dense $p=1$ cases are illustrated in Figure~\ref{fig:p1_vs_p3_3_cyl}, including data from Yu et al \cite{Yu2016FlowInduced}.
For the upstream cylinder, both candidates produce comparable results with a slight $x-$offset for $p=3$ of the order $\delta x^* \approx 0.02$.
For the second cylinder, similar trajectories are observed for both cases, although the $p=1$ solution begins to show slight deviations. These discrepancies become more pronounced for the downstream cylinder, where the lower-order solution departs significantly from the reference trajectories, while the $p=3$ case remains in good agreement.

The wake structures are further compared through vorticity contours for the two discretization orders. Overall, both $p=1$ and $p=3$ simulations reproduce similar close-body wake patterns and larger-scale vortex topology, e.g., the same ``2P'' pattern shed from the downstream cylinder's boundary and a similar ``second-wake'' image. However, notable differences are observed at approximately $5D$ distance downstream of the last cylinder.

\begin{figure}[h]
	\centering
	\includegraphics[width=0.85\textwidth,keepaspectratio]{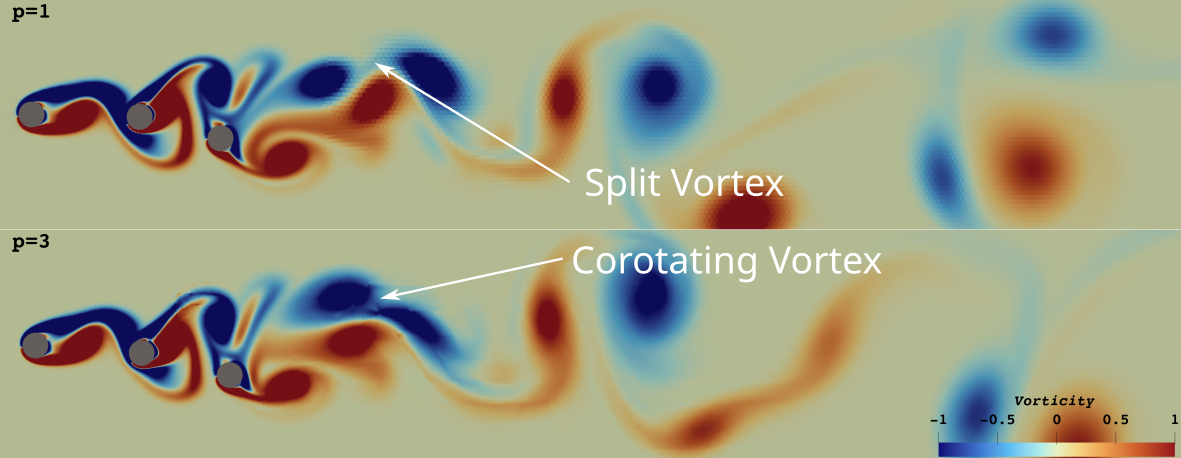}
	\caption{Vortex wake comparison between the two orders at $\sim 356 s$ for the reduced velocity of $U^*=10$.}
	\label{fig:p1_vs_p3_3_cyl_vort_tm1}
\end{figure}

In detail, at approximately $t=356\ s$, as shown in Figure \ref{fig:p1_vs_p3_3_cyl_vort_tm1}, for $p=1$ two co-rotating vortices are split and a contra-rotating vortex intercepts, while for $p=3$ these exact two vortices are tightly merged and the contra-rotating vortex is advected parallel to the flow, which affects the third cylinder's oscillation.

This illustrates how the higher-order discretization better preserves the physical characteristics of the wake, particularly in regions of strong vortex interaction and pairing. In contrast, the near-body flow field remains qualitatively similar in both simulations, which explains the closer agreement in the trajectories of the front and middle cylinders.

Having established the accuracy differences, we now examine the computational cost of each discretization strategy. First, we consider the total number of DoFs of the system, $N_{\mathrm{dof}} = N_E N_p$, where $N_E$ is the number of elements and $N_p$ is the number of polynomial degrees of freedom per element. Also, $N_T$ is the total number of explicit time iterations required for each simulation to reach $600\ s$; $N_T$ is dictated by the CFL condition.
Computing cost can be ``loosely'' summarized as the product $N_T N_{\mathrm{dof}}$, combining, thus, a measure of both spatial and temporal computational effort. The resulting data are shown in Table~\ref{tab:p1_vs_p3_comp} where we observe that the first-order simulation is more costly than the third-order simulation. Lastly, as the results suggest, the $p=1$ case would require an even finer grid to match the accuracy of the higher-order solution, making the $p=3$ alternative a far more efficient choice.

\begin{table}[h]
	\centering
	\renewcommand{\arraystretch}{1.3}  
	\begin{tabular}{c|ccccc}
		\hline
		$p$ & $N_p$ & $N_E$    & $N_{\mathrm{dof}}$ & $N_T$     & $N_T N_{\mathrm{dof}}$     \\
		\hline
		$1$ & $3 $  & $279212$ & $837636$           & $1188500$ & $\sim 9.955 \times10^{11}$ \\
		$3$ & $10$  & $14707 $ & $147070$           & $4715000$ & $\sim 6.934 \times10^{11}$ \\
		\hline
	\end{tabular}
	\caption{Total degrees of freedom and time iterations for each polynomial order.}
	\label{tab:p1_vs_p3_comp}
\end{table}

%% file: conclusions3.tex
In this work, we developed a high-order ALE DG framework for simulating highly nonlinear multi-body VIV phenomena. The framework extends a scalable Runge-Kutta Interior-Penalty Discontinuous Galerkin (RK IPDG) compressible CFD solver to moving domains, while the use of affine triangular elements ensures that the additional overhead due to ALE remains minimal. The Geometric Conservation Law (GCL) is enforced by numerically marching the Jacobian's determinant alongside the main system, ensuring free-stream preservation up to machine precision on deforming meshes. Mesh deformation is handled using a weighted Radial Basis Function technique, enabling the simulation of extreme cases of large structural displacements encountered in this work.

The framework was applied to two progressively complex configurations. The first case, a two-cylinder tandem arrangement with one DoF, showed excellent agreement with the results of both Griffith et al. \cite{griffith2017flow} and Papadakis et al. \cite{papadakis2022hybrid} across the Lissajous curves, oscillation power spectra, and vortex shedding modes. A key observation is the preservation of wake structures over long distances even on a relatively coarse mesh, highlighting the low numerical diffusion feature of the method.

The second case, a three-cylinder tandem arrangement with two DoF, presented highly irregular trajectories driven by complex multi-body wake interactions. Despite this complexity, the results are in good agreement with the data of Yu et al.\cite{Yu2016FlowInduced}. Poincaré phase plots confirmed that the cylinders' motions deviate substantially from classical harmonic VIV responses. A notable feature is the periodic ``attract-and-release'' cycle of the trailing cylinder at higher reduced velocities, where alternating wake states periodically suppress and amplify the stream-wise oscillations, in close agreement with the findings 
of \cite{Yu2016FlowInduced}.

Finally, an $hp$-refinement study on the three-cylinder case at $U^*=10$ demonstrated that $p$-refinement is more efficient than $h$-refinement. The higher-order discretization ($p=3$) better preserved both the cylinder trajectories and the wake structures compared to a significantly denser $p=1$ mesh, while also being computationally less expensive. This underscores the advantage of high-order methods for FSI applications where the accurate resolution of vortex dynamics is critical for predicting the structural response.


Concluding, this work demonstrates that an ALE high-order DG framework can effectively handle the complex dynamics of multi-body VIV problems on relatively coarse meshes, offering a promising alternative to traditional low-order methods that require finer grids to achieve similar accuracy. Given that the DG methodology naturally extends to three dimensions and that an implicit time integration strategy can be readily incorporated into the current framework, future work will target three-dimensional simulations at higher Reynolds numbers, representative of more realistic engineering conditions.

%% file: appendix.tex
In this section a proof of \eqref{eq:gcl} is given. 
Starting with \eqref{eq:freestream}, the term multiplied with $\bar{\vect{U}}$, after mapping back to the physical coordinates $\vect{x}$ reads,
\begin{equation}
    \frac{d}{dt} \int_{\mathcal{D}^K (t) } \psi_n^K \Idiffx
    + \int_{\mathcal{D}^K (t) } \mathbf{v}^K \cdot  \nabla \psi_n^K \Idiffx
    - \int_{\partial \mathcal{D}^K (t)} \vect{n}^K \cdot \mathbf{v}^K \, \psi_n^K \Idiffx = 0.
    \label{eq:appendx_GCL1}
\end{equation}
The application of the divergence theorem on the last term of \eqref{eq:appendx_GCL1} yields,
\begin{equation}
    \int_{\partial \mathcal{D}^K (t)} \vect{n}^K \cdot \mathbf{v}^K \, \psi_n^K \Idiffx = 
    \int_{\mathcal{D}^K (t)} \nabla \cdot ( \mathbf{v}^K \, \psi_n^K ) \Idiffx.
    \label{eq:appendx_GCL2}
\end{equation}
Now after substituting \eqref{eq:appendx_GCL2} on \eqref{eq:appendx_GCL1}, we have,
\begin{equation}
    \frac{d}{dt} \int_{\mathcal{D}^K (t)} \psi_n^K \Idiffx + \int_{\mathcal{D}^K (t)} \mathbf{v}^K \cdot \nabla \psi_n^K - \nabla \cdot ( \mathbf{v}^K \, \psi_n^K) \Idiffx = 0. \label{eq:appendx_GCL3}
\end{equation}
Using now the identity $\nabla \cdot (\vect{a} b) = b \nabla \cdot \vect{a} + \vect{a} \cdot \nabla b$, \eqref{eq:appendx_GCL3} can be written as,
\begin{equation}
    \frac{d}{dt} \int_{\mathcal{D}^K (t)} \psi_n^K \Idiffx - \int_{\mathcal{D}^K (t)} \psi_n^K \nabla  \cdot\mathbf{v}^K  \Idiffx = 0. \label{eq:appendx_GCL4}
\end{equation}
Before we map onto the reference space, we need the following identity for the divergence operator, 
$\nabla \cdot \mathbf{v}^K = \nabla_{\vect{r}} \cdot \big( \mathcal{J}^K  ( \mathcal{G}^K_{\vect{r}} )^{-1} \mathbf{v}^K \big) / \mathcal{J}^K $,  
which due to the constant Jacobian in space, this relation simplifies to 
$\nabla \cdot \mathbf{v}^K = \nabla_{\vect{r}} \cdot \big( ( \mathcal{G}^K_{\vect{r}} )^{-1} \mathbf{v}^K \big) $.

Then, \eqref{eq:appendx_GCL4} reads,
\begin{equation}
     \int_{\mathcal{D}} \psi_n \Big( \frac{d}{dt} \mathcal{J}^K - \mathcal{J}^K \nabla_{\vect{r}} \cdot \big( ( \mathcal{G}^K_{\vect{r}} )^{-1} \mathbf{v}^K \big) \Big)  \Idiffr = 0.  \label{eq:appendx_GCL5}
\end{equation}

Since \eqref{eq:appendx_GCL5} holds for all $\psi_n(\vect{r}) \in  \mathbb{P}_K$, 
equation \eqref{eq:gcl} follows directly from \eqref{eq:appendx_GCL5}. 

Also, after using simple algebra, the exact form of the constant $C^K$ is, 
\begin{equation}
    C^K = \frac{1}{2} \left( r_x^K (v_{2,x}^K - v_{1,x}^K ) + r_y^K (v_{2,y}^K - v_{1,y}^K ) +
    s_x^K (v_{3,x}^K - v_{1,x}^K ) + s_y^K (v_{3,y}^K - v_{1,y}^K ) \right). 
    \label{GCL_eq}
\end{equation}
With $\mathbf{x}^K_i (t) = \big[x_{i}^K,  y_{i}^K \big]^T$, $ \mathbf{v}^K_i (t) = \big[v_{i,x}^K,  v_{i,y}^K \big]^T$ and $\mathbf{v}^K_i = (d/dt)\mathbf{x}^K_i$ for $i = 1,2,3$. Lastly, the inverse of the Jacobian matrix is,
\begin{equation}
\renewcommand{\arraystretch}{1.3} 
\big( \mathcal{G}^K_{\vect{r}} \big)^{-1} =
\left( \begin{array}{cc}
   r_x^K  & r_y^K \\
   s_x^K  & s_y^K
\end{array} \right) = 
\frac{1}{2 \mathcal{J}^K} \left( \begin{array}{cc}
        y_{3}^K - y_{1}^K   & -(y_{2}^K - y_{1}^K) \\
       -(x_{3}^K - x_{1}^K) & x_{2}^K - x_{1}^K
    \end{array} \right)
\label{eq:appendx_diff_G}
\end{equation}
\renewcommand{\arraystretch}{1} 